\documentclass[twocolumn]{aastex63}
\usepackage{inputenc}
\usepackage{comment}
\usepackage{colortbl}
\usepackage{graphicx}
\usepackage{soul}
\usepackage{hyperref}
\usepackage{multirow}
\usepackage{amsmath}

\shorttitle{Multi-band Analysis of J2027}

\graphicspath{{./}{figures/}}
\begin{document}

\title{HAWC, Fermi-LAT and Chandra Detections of High Energy Particles in the S106 Star Forming Region}

\author{R.~Alfaro}
\affiliation{Instituto de F\'{i}sica, Universidad Nacional Autónoma de México, Ciudad de Mexico, Mexico}
\author{C.~Alvarez}
\affiliation{Universidad Autónoma de Chiapas, Tuxtla Gutiérrez, Chiapas, México}

\author[0000-0002-0595-9267]{M.~Araya}
\affiliation{Universidad de Costa Rica, San José 2060, Costa Rica}
\author{J.C.~Arteaga-Velázquez}
\affiliation{Universidad Michoacana de San Nicolás de Hidalgo, Morelia, Mexico}
\author[0000-0002-4020-4142]{D.~Avila Rojas}
\affiliation{Instituto de Astronom\'{i}a, Universidad Nacional Autónoma de México, Ciudad de México, México.}

\author[0000-0002-2084-5049]{H.A.~Ayala Solares}
\affiliation{Temple University, Department of Physics, 1925 N. 12th Street, Philadelphia, PA 19122, USA}

\author[0000-0002-5529-6780]{R.~Babu}
\affiliation{Department of Physics and Astronomy, Michigan State University, East Lansing, MI, USA}
\author{P.~Bangale}
\affiliation{Temple University, Department of Physics, 1925 N. 12th Street, Philadelphia, PA 19122, USA}
\author{A.~Bernal}
\affiliation{Instituto de Astronom\'{i}a, Universidad Nacional Autónoma de México, Ciudad de México, México.}
\author[0000-0002-4042-3855]{K.S.~Caballero-Mora}
\affiliation{Universidad Autónoma de Chiapas, Tuxtla Gutiérrez, Chiapas, México}
\author[0000-0003-2158-2292]{T.~Capistrán}
\affiliation{INFN Sezione di Torino, Via P. Giuria 1, 10125 Torino, Italy}

\author[0000-0002-8553-3302]{A.~Carramiñana}
\affiliation{Instituto Nacional de Astrof\'{i}sica, Óptica y Electrónica, Puebla, Mexico}
\author[0000-0002-6144-9122]{S.~Casanova}
\affiliation{Institute of Nuclear Physics Polish Academy of Sciences, PL-31342 IFJ-PAN, Krakow, Poland}
\author{A.L.~Colmenero-Cesar}
\affiliation{Universidad Michoacana de San Nicolás de Hidalgo, Morelia, Mexico}

\author[0000-0002-7607-9582]{U.~Cotti}
\affiliation{Universidad Michoacana de San Nicolás de Hidalgo, Morelia, Mexico}
\author[0000-0002-1132-871X]{J.~Cotzomi}
\affiliation{Facultad de Ciencias F\'{i}sico Matemáticas, Benemérita Universidad Autónoma de Puebla, Puebla, Mexico}
\author[0000-0002-7574-1298]{N.~Di Lalla}
\affiliation{Department of Physics, Stanford University: Stanford, CA 94305–4060, USA}
\author{R.~Diaz Hernandez}
\affiliation{Instituto Nacional de Astrof\'{i}sica, Óptica y Electrónica, Puebla, Mexico}
\author[0000-0002-2987-9691]{M.A.~DuVernois}
\affiliation{Dept. of Physics and Wisconsin IceCube Particle Astrophysics Center, University of Wisconsin{\textemdash}Madison, Madison, WI, USA}
\author[0000-0001-5737-1820]{K.~Engel}
\affiliation{Department of Physics, University of Maryland, College Park, MD, USA}
\author[0000-0003-2423-4656]{T.~Ergin}
\affiliation{Department of Physics and Astronomy, Michigan State University, East Lansing, MI, USA}
\author[0000-0001-7074-1726]{C.~Espinoza}
\affiliation{Instituto de F\'{i}sica, Universidad Nacional Autónoma de México, Ciudad de Mexico, Mexico}
\author[0000-0002-0173-6453]{N.~Fraija}
\affiliation{Instituto de Astronom\'{i}a, Universidad Nacional Autónoma de México, Ciudad de México, México.}
\author{S.~Fraija}
\affiliation{Instituto de Astronom\'{i}a, Universidad Nacional Autónoma de México, Ciudad de México, México.}

\author[0000-0002-4188-5584]{J.A.~García-González}
\affiliation{Tecnologico de Monterrey, Escuela de Ingenier\'{i}a y Ciencias, Ave. Eugenio Garza Sada 2501, Monterrey, N.L., Mexico, 64849}
\author[0000-0003-1122-4168]{F.~Garfias}
\affiliation{Instituto de Astronom\'{i}a, Universidad Nacional Autónoma de México, Ciudad de México, México.}
\author{N.~Ghosh}
\affiliation{Department of Physics, Michigan Technological University, Houghton, MI, USA}
\author{S.~Giri}
\affiliation{Institute of Nuclear Physics Polish Academy of Sciences, PL-31342 IFJ-PAN, Krakow, Poland}
\author{A.~Gonzalez Muñoz}
\affiliation{Instituto de F\'{i}sica, Universidad Nacional Autónoma de México, Ciudad de Mexico, Mexico}
\author[0000-0002-5209-5641]{M.M.~González}
\affiliation{Instituto de Astronom\'{i}a, Universidad Nacional Autónoma de México, Ciudad de México, México.}
\author{J.A.~González}
\affiliation{Universidad Michoacana de San Nicolás de Hidalgo, Morelia, Mexico}
\author[0000-0002-9790-1299]{J.A.~Goodman}
\affiliation{Department of Physics, University of Maryland, College Park, MD, USA}
\author{S.~Groetsch}
\affiliation{Department of Physics, Michigan Technological University, Houghton, MI, USA}
\author{J.~Gyeong}
\affiliation{Department of Physics, Sungkyunkwan University, Suwon 16419, South Korea}
\author[0000-0001-9844-2648]{J.P.~Harding}
\affiliation{Los Alamos National Laboratory, Los Alamos, NM, USA}
\author[0000-0001-5169-723X]{I.~Herzog}
\affiliation{Department of Physics and Astronomy, Michigan State University, East Lansing, MI, USA}
\author[0000-0002-5447-1786]{D.~Huang}
\affiliation{Department of Physics and Astronomy, University of Delaware, Newark, DE, USA}
\author[0000-0002-3302-7897]{P.~Hüntemeyer}
\affiliation{Department of Physics, Michigan Technological University, Houghton, MI, USA}
\author{S.~Kaufmann}
\affiliation{Universidad Politecnica de Pachuca, Pachuca, Hgo, Mexico}
\author[0000-0003-4785-0101]{D.~Kieda}
\affiliation{Department of Physics and Astronomy, University of Utah, Salt Lake City, UT, USA}
\author[0000-0001-6336-5291]{A.~Lara}
\affiliation{Instituto de Geof\'{i}sica, Universidad Nacional Autónoma de México, Ciudad de Mexico, Mexico}

\author{K.~Leavitt}
\affiliation{Department of Physics, Michigan Technological University, Houghton, MI, USA}
\author[0000-0001-5516-4975]{H.~León Vargas}
\affiliation{Instituto de F\'{i}sica, Universidad Nacional Autónoma de México, Ciudad de Mexico, Mexico}
\author[0000-0003-2696-947X]{J.T.~Linnemann}
\affiliation{Department of Physics and Astronomy, Michigan State University, East Lansing, MI, USA}

\author[0000-0001-8825-3624]{A.L.~Longinotti}
\affiliation{Instituto de Astronom\'{i}a, Universidad Nacional Autónoma de México, Ciudad de México, México.}
\author[0000-0003-2810-4867]{G.~Luis-Raya}
\affiliation{Universidad Politecnica de Pachuca, Pachuca, Hgo, Mexico}
\author[0000-0001-8088-400X]{K.~Malone}
\affiliation{Los Alamos National Laboratory, Los Alamos, NM, USA}
\author[0000-0001-9052-856X]{O.~Martinez}
\affiliation{Facultad de Ciencias F\'{i}sico Matemáticas, Benemérita Universidad Autónoma de Puebla, Puebla, Mexico}
\author[0000-0002-2824-3544]{J.~Martínez-Castro}
\affiliation{Centro de Investigaci\'on en Computaci\'on, Instituto Polit\'ecnico Nacional, M\'exico City, M\'exico.}
\author[0000-0002-2610-863X]{J.A.~Matthews}
\affiliation{Dept of Physics and Astronomy, University of New Mexico, Albuquerque, NM, USA}
\author[0000-0002-8390-9011]{P.~Miranda-Romagnoli}
\affiliation{Universidad Autónoma del Estado de Hidalgo, Pachuca, Mexico}
\author{P.E.~Mirón-Enriquez}
\affiliation{Instituto de Astronom\'{i}a, Universidad Nacional Autónoma de México, Ciudad de México, México.}
\author[0000-0002-1114-2640]{E.~Moreno}
\affiliation{Facultad de Ciencias F\'{i}sico Matemáticas, Benemérita Universidad Autónoma de Puebla, Puebla, Mexico}
\author[0000-0002-7675-4656]{M.~Mostafá}
\affiliation{School of Science \& Engineering, University of Missouri--Kansas City, Kansas City, MO, USA}
\author{M.~Najafi}
\affiliation{Department of Physics, Michigan Technological University, Houghton, MI, USA}
\author[0000-0003-1059-8731]{L.~Nellen}
\affiliation{Instituto de Ciencias Nucleares, Universidad Nacional Autónoma de Mexico, Ciudad de Mexico, Mexico}
\author[0000-0002-6859-3944]{M.U.~Nisa}
\affiliation{Department of Physics and Astronomy, Michigan State University, East Lansing, MI, USA}
\author[0000-0001-7099-108X]{R.~Noriega-Papaqui}
\affiliation{Universidad Autónoma del Estado de Hidalgo, Pachuca, Mexico}
\author[0000-0002-5448-7577]{N.~Omodei}
\affiliation{Department of Physics, Stanford University: Stanford, CA 94305–4060, USA}

\author[0009-0009-2481-6921]{M.~Osorio-Archila}
\affiliation{Instituto de Astronom\'{i}a, Universidad Nacional Autónoma de México, Ciudad de México, México.}
\author{E.~Ponce}
\affiliation{Facultad de Ciencias F\'{i}sico Matemáticas, Benemérita Universidad Autónoma de Puebla, Puebla, Mexico}
\author[0000-0001-5998-4938]{E.G.~Pérez-Pérez}
\affiliation{Universidad Politecnica de Pachuca, Pachuca, Hgo, Mexico}
\author[0000-0002-6524-9769]{C.D.~Rho}
\affiliation{Department of Physics, Sungkyunkwan University, Suwon 16419, South Korea}

\author[0000-0003-1327-0838]{D.~Rosa-González}
\affiliation{Instituto Nacional de Astrof\'{i}sica, Óptica y Electrónica, Puebla, Mexico}
\author{D.~Salazar-Gallegos}
\affiliation{Department of Physics and Astronomy, Michigan State University, East Lansing, MI, USA}

\author[0000-0001-6079-2722]{A.~Sandoval}
\affiliation{Instituto de F\'{i}sica, Universidad Nacional Autónoma de México, Ciudad de Mexico, Mexico}
\author[0000-0001-8644-4734]{M.~Schneider}
\affiliation{Department of Physics, University of Maryland, College Park, MD, USA}
\author{M.~Shin}
\affiliation{Department of Physics, Sungkyunkwan University, Suwon 16419, South Korea}

\author[0000-0002-1012-0431]{A.J.~Smith}
\affiliation{Department of Physics, University of Maryland, College Park, MD, USA}
\author{Y.~Son}
\affiliation{Department of Physics, Sungkyunkwan University, Suwon 16419, South Korea}
\author[0000-0002-1492-0380]{R.W.~Springer}
\affiliation{Department of Physics and Astronomy, University of Utah, Salt Lake City, UT, USA}
\author{O.~Tibolla}
\affiliation{Universidad Politecnica de Pachuca, Pachuca, Hgo, Mexico}
\author[0000-0001-9725-1479]{K.~Tollefson}
\affiliation{Department of Physics and Astronomy, Michigan State University, East Lansing, MI, USA}
\author[0000-0002-1689-3945]{I.~Torres}
\affiliation{Instituto Nacional de Astrof\'{i}sica, Óptica y Electrónica, Puebla, Mexico}
\author[0000-0002-7102-3352]{R.~Torres-Escobedo}
\affiliation{Tsung-Dao Lee Institute \& School of Physics and Astronomy, Shanghai Jiao Tong University, 800 Dongchuan Rd, Shanghai, SH 200240, China}
\author[0000-0001-6798-353X]{X.~Wang}
\affiliation{Department of Physics, Missouri University of Science and Technology, Rolla, MO, US}
\author{Z.~Wang}
\affiliation{Department of Physics, Missouri University of Science and Technology, Rolla, MO, US}
\author[0000-0003-2141-3413]{I.J.~Watson}
\affiliation{University of Seoul, Seoul, Rep. of Korea}
\author[0009-0005-7243-1402]{H.~Wu}
\affiliation{Dept. of Physics and Wisconsin IceCube Particle Astrophysics Center, University of Wisconsin{\textemdash}Madison, Madison, WI, USA}
\author[0009-0006-3520-3993]{S.~Yu}
\affiliation{Department of Physics, Pennsylvania State University, University Park, PA, USA}
\author[0000-0002-9307-0133]{S.~Yun-Cárcamo}
\affiliation{Dept. of Physics, Drexel University, 3141 Chestnut Street, Philadelphia, PA 19104, USA}
\author{X.~Zhang}
\affiliation{Institute of Nuclear Physics Polish Academy of Sciences, PL-31342 IFJ-PAN, Krakow, Poland}
\author[0000-0002-8528-9573]{C.~de León}
\affiliation{Universidad Michoacana de San Nicolás de Hidalgo, Morelia, Mexico}
\collaboration{83}{(HAWC Collaboration)}
% Steve
\author[0000-0003-2633-2196]{Stephen DiKerby}
\affiliation{Department of Physics and Astronomy, Michigan State University, East Lansing, MI, USA}

% Shuo
\author[0000-0002-2967-790X]{Shuo Zhang}
\affiliation{Department of Physics and Astronomy, Michigan State University, East Lansing, MI, USA}

\correspondingauthor{R.~Babu, S.~DiKerby, T.~Ergin, S.~Zhang, M.U.~Nisa}
\email{baburish@msu.edu, dikerbys@msu.edu, tulun.ergin@wright.edu, zhan2214@msu.edu, nisamehr@msu.edu}
%%%% Abstract
\begin{abstract}

Local star-forming regions have recently been the subject of scrutiny as cosmic-ray accelerators and sources of ultra-high-energy gamma-ray emissions. The Cygnus-X region is one of the richest star forming regions in the galaxy and over the last few decades has been extensively studied. Here we report the detection of a star-forming region located in the southern part of Cygnus-X, Sharpless 106 (S106), using the HAWC Observatory. The S106 region is spatially located near the cataloged gamma-ray source 1LHAASO J2027+3657. In this work, we perform a multi-wavelength analysis of S106, by combining HAWC data with re-analysis of \textit{Fermi}-LAT and archival \textit{Chandra} data. We construct a fully leptonic model for the multi-wavelength emission from S106 from a single population of high-energy leptons, with the gamma-rays greater than 1 TeV being produced by inverse Compton scattering of the locally-produced IR photon field from S106 itself. Our models show that hadronic mechanisms are not always required to explain the observed flux from gamma-ray star-forming regions and that understanding the specific details of local photon fields is crucial to modeling the multi-wavelength emission from star-forming regions. 

\end{abstract}

%%%% Keywords

%%%% Introduction and Context
\section{Introduction}
\label{sec:Intro}

Recent surveys by Very High Energy (VHE; 100 GeV–100 TeV) and Ultra High Energy (UHE; $>100$ TeV) instruments like the High-Altitude Water Cherenkov Observatory (HAWC) \citep{2020ApJ...905...76A} and the Large High Altitude Air Shower Observatory (LHAASO) \citep{2024ApJS..271...25C} have discovered a substantial population of TeV and PeV sources concentrated in the galactic plane, but the nature of these objects is still uncertain. Given that these gamma-ray sources emit particles and photons at energies far exceeding those in human-made particle accelerators, modeling the processes that produce their gamma-ray emission can contribute to the identification of new astrophysical laboratories to test the limits of our fundamental theories.

Several ongoing surveys are conducting low-energy investigations of promising galactic gamma-ray sources that lack confident association with known counterparts. Studies over the past decade have revealed that galactic gamma-ray sources can be attributed to a wide range of astronomical phenomena, including pulsars, pulsar wind nebulae (PWN), and supernova remnants (SNR). Despite wide-ranging interest, many gamma-ray sources remain ``unassociated'', lacking confident low-energy counterparts. Identifying the low-energy counterparts of unassociated gamma-ray sources is a vital step towards completing the census of extreme systems within our own galaxy.

Whether high-energy systems produce observed gamma rays via leptonic -- synchrotron, inverse Compton (IC), Bremsstrahlung -- or hadronic -- pion decay, proton synchrotron -- processes is a particularly timely question, given the recent discovery of high-energy neutrino fluxes from flaring blazars \citep{2018Sci...361..147I}, the galactic plane \citep{2023Sci...380.1338I}, and the nearby Seyfert galaxy NGC 1068 \citep{2022Sci...378..538I}. High-energy neutrino emission requires hadronic processes that produce or destroy leptons, creating neutrinos to conserve lepton number. The identification of galactic gamma-ray emitters as hadronic can help explain the observed galactic neutrino flux.

Giant molecular clouds are the birthplaces of stars and hosts to young stellar objects. These star-forming regions (SFRs) often display strong infrared emission, non-thermal radio signatures, and maser activity, which serve as key tracers of on-going star formation. The production of gamma-rays in such environments can occur due to interactions between accretion processes, collimated jets, stellar winds, and ambient material \citep{Ro10,Er20}. In these ways, interactions between cluster-driven cosmic rays and surrounding dense gas can produce diffuse gamma-ray emission \citep{Ac11,Ab12,Ah19,Sa20}.

Besides high-energy processes related to first-order Fermi acceleration, efficient particle acceleration can also produce non-thermal radiation from radio to X-ray energies. The power distributed into a SFR from even a single young stellar system is considerable; in massive star binaries, a substantial fraction of the wind kinetic power (exceeding $\sim10^{33}\:\rm{erg/s}$) can be transferred to high-energy particles and detectable gamma-ray emission \citep{Ta09,Be11,Re15,Ps16}. In addition, protons might escape the system, interacting with nearby ambient matter. The collective activities of massive stars may form super-bubbles (e.g. Cygnus OB2 or the Cygnus Cocoon), where the acceleration of multiple shocks can raise the maximum energy of CR protons above $1 \:\rm{PeV}$ \citep{Ah19,LHCygnusC2024}.
 
The gamma-ray source 1LHAASO J2027+3657 (henceforth J2027) is an unassociated source \citep{2024ApJS..271...25C} detected in the UHE KM2A component of LHAASO but not by the WCDA instrument. J2027's position in the outer galactic plane, its moderate flux ($N_0 = (5.0 \pm 0.6) \times 10^{-17} \:\rm{/s/cm^2/TeV}$ and $E_0 = 50\ \rm{TeV}$) but steep power-law slope ($\Gamma = 3.21 \pm 0.17$), and lack of counterparts make it a prime target for low-energy follow-up and analysis. In particular, determining whether the gamma-ray emission from J2027 is leptonic, hadronic, or a combination of both is a crucial step in evaluating its potential contribution to the galactic plane neutrino flux observed in IceCube \citep{2023Sci...380.1338I}.

A possible counterpart to J2027 is the nearby bipolar star-forming and HII region Sharpless 106 (henceforth S106). It is located at a distance of $\sim 1.4 \:\rm{kpc}$ \citep{adams2015}. \cite{Comeron_2018} revealed that the central object in the region, S106IR, is a close binary system, possibly responsible for the two-lobed structure of S106 as a whole. S106 also hosts a collection of young stars, many of which are strong X-ray emitters previously studied in \cite{2004A&A...424..965G}, and there may be diffuse X-ray flux from the collective interacting stellar winds of stars in S106.
 
The \textit{Fermi}-LAT source catalog (4FGL-DR4, \cite{4FGLDR4}) includes the point-like gamma-ray source 4FGL J2026.5+3718c, which positionally coincides with S106, but is $\sim 20'$ away from the J2027's position as measured by the LHAASO-KM2A. Re-analysis of this archival \textit{Fermi}-LAT data, along with the HAWC data analysis presented in this work, will more tightly constrain the gamma-ray landscape around S106 beyond the spectral point provided in the LHAASO catalog. 

In this work we report analysis of a range of high-energy data to build a multiwavelength spectrum near J2027, testing what connection J2027 might have with S106 and other gamma-ray sources detected in \textit{Fermi}-LAT and HAWC. In section \ref{sec:ObsAna} we describe the analysis of HAWC, \textit{Fermi}-LAT, and Chandra data to construct a broadband SED for J2027. In section \ref{sec:MWLMod} we discuss several multiwavelength models linking the observed electromagnetic emission to underlying particle populations. Finally, in section \ref{sec:DiscConc} we present overall conclusions.

%%%% Observations and Data Analysis
\section{Observations and Data Analysis}
\label{sec:ObsAna}

%%% HAWC Analysis 
\subsection{HAWC Analysis}
\label{sec:HAWCAna}

%% FIGURE 1
\begin{figure}
\centering \vspace*{1pt}
\includegraphics[width=0.5\textwidth]{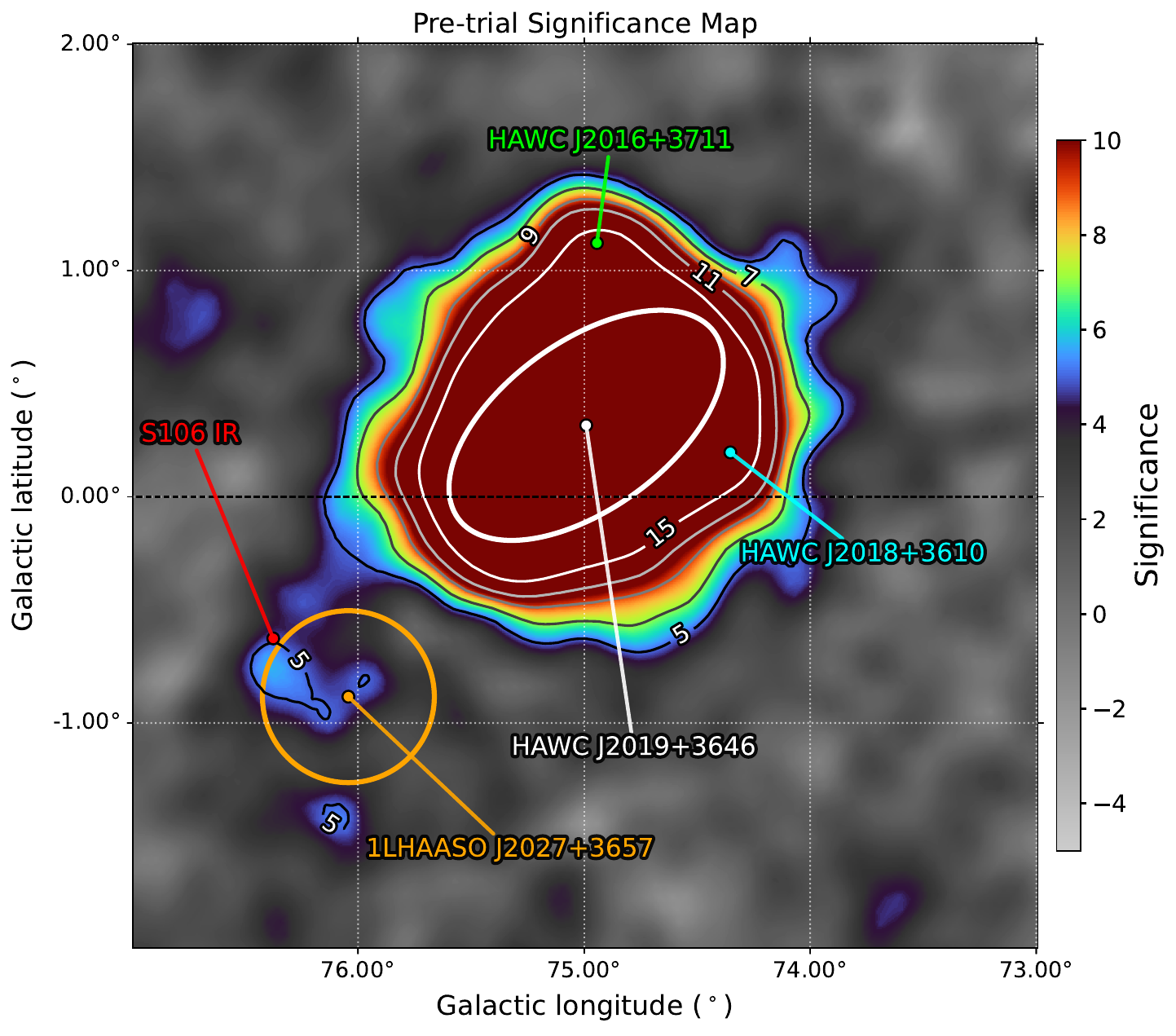}
\caption{ {\footnotesize HAWC pre-trial significance Map of the region created using the NN energy estimator map including off-array showers. This map is created assuming a point-like source morphology and a power-law type spectrum having a spectral index of 2.7. The galactic plane (b=0$^\circ$) is shown in gray dashed lines. The orange label shows the LHAASO location of J2027. }}
\label{fig:HAWC_OnOff_NNMap}
\end{figure}

%% FIGURE 2 
\begin{figure*}
\centering \vspace*{1pt}
\includegraphics[width=0.4\textwidth]{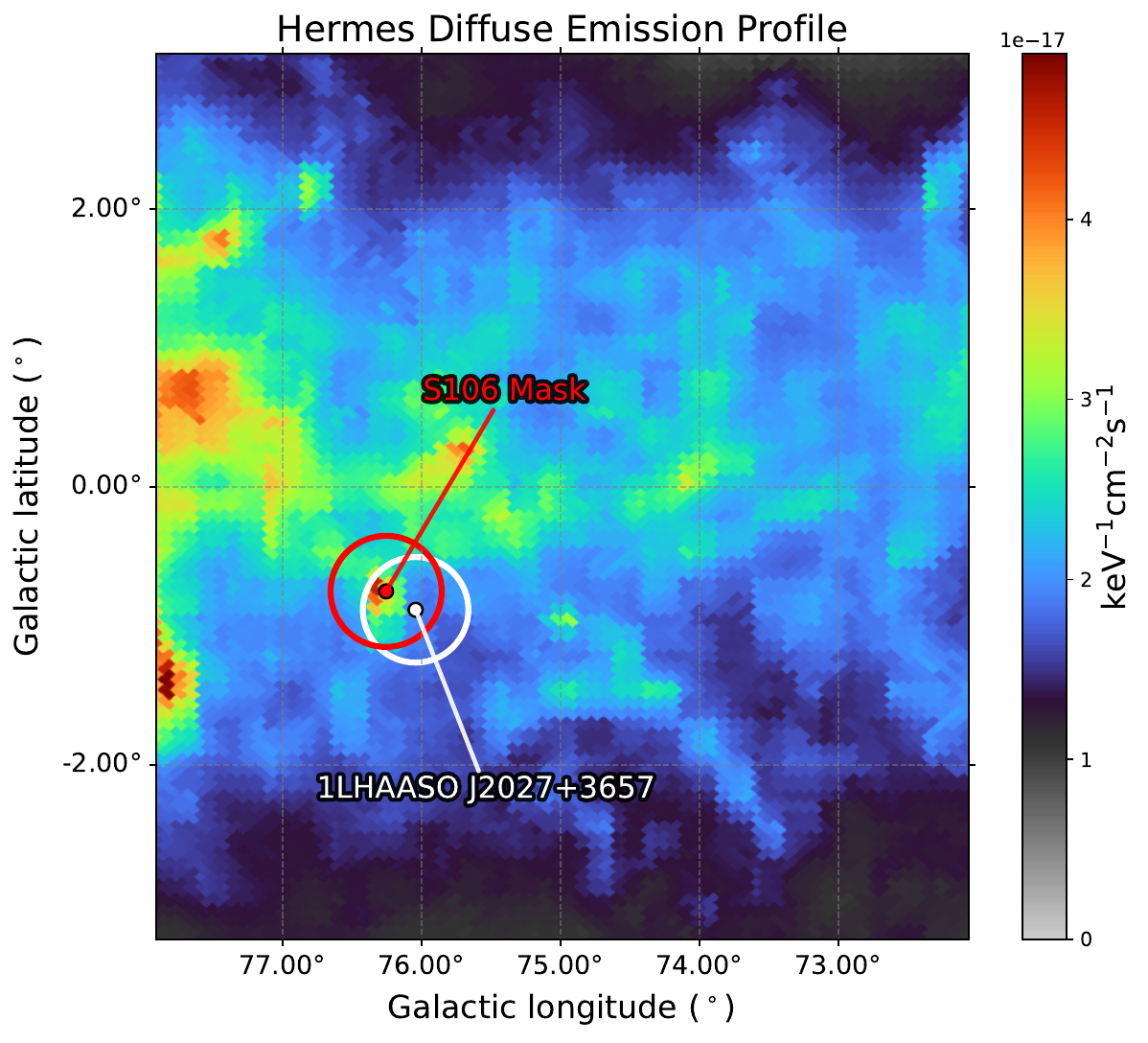}
\includegraphics[width=0.45\textwidth]{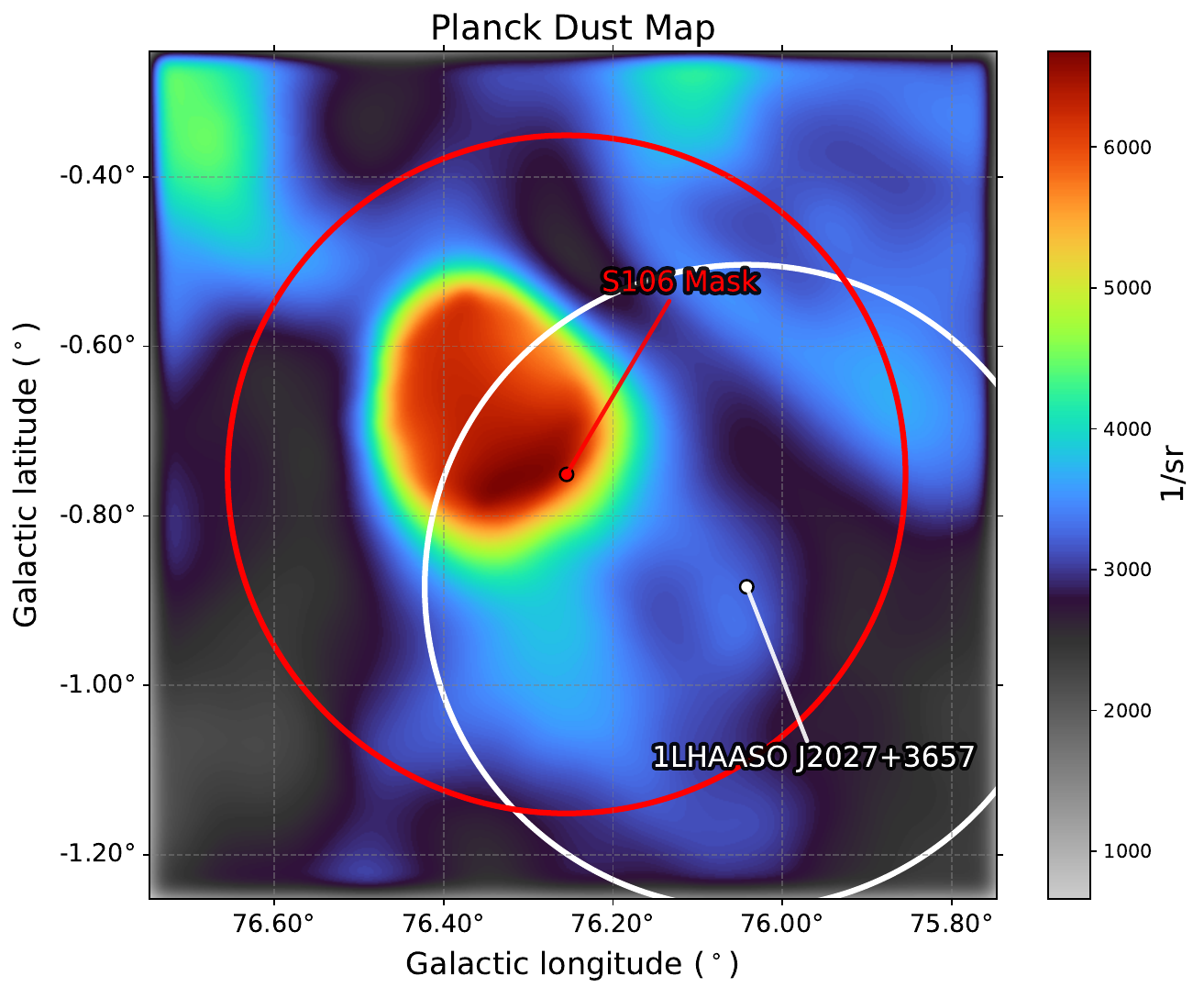}
\caption{ {{\footnotesize {\it Left Panel:} The HERMES diffuse emission profile templates used in this analysis. The HERMES template for the ICS+$\pi^0$(HI+H$_2$) emission profile along with the location of 1LHAASO J2027+3657 (white) and the 0.5$^\circ$ mask at the centroid of the peak $^{12}$CO emission profile (red).  {\it Right Panel:} Planck dust opacity map of the region surrounding the dense $^{12}$CO distribution spatially overlapping with S106's position.} }}
\label{fig:DiffEmissionMaps}
\end{figure*}

%% HAWC Data and Analysis Setup
\subsubsection{ HAWC Data and Analysis Setup}
\label{HAWCData}
To search for emission from J2027 in HAWC, we analyze $\sim2900$ days of HAWC Pass 5 data via the neural network (NN) energy estimator above 1 TeV. This dataset includes on-array events, as well as off-array events with reconstructed shower cores falling off the main array up to 1.5 times its physical area, as discussed in \citet{HAWCCrab2019} and \citet{HAWCP5}. This dataset has stricter gamma-ray/hadron separation cuts based on a machine-learning algorithm, which further improves our sensitivity at lower energies \citep{HAWCML}. 

We perform a binned maximum likelihood analysis of the data using the \texttt{threeML} \citep{threeML} with HAWC Accelerated Likelihood (HAL) plugin \texttt{hawc-hal} \citep{hawchal}. To quantify the goodness of fit of the sources in the region, we use a test statistic (TS), defined as:
\begin{equation}
    TS = 2 \times (\mathrm{ln}\ L_1 - \mathrm{ln}\ L_0 )
\end{equation}
\noindent  where $L_1$ and $L_0$ are the likelihoods of the presence (alternate hypothesis) and absence (null hypothesis) of a source respectively. Following Wilks' theorem, we express the pre-trial significance as $\sqrt{TS}$. We use the source search pipeline outlined in \citet{hawcic443}. In this work, we enhance the point source search stage by integrating the image processing driven source seed finder introduced in \citet{hawcimage}. %This improvement enhances pipeline performance, yielding an accurate model for the ROI with substantially reduced computational time. 

The HAWC pre-trial significance at the location of J2027 is $\sim 6.6 \sigma$ in the NN energy estimator map as shown in Figure \ref{fig:HAWC_OnOff_NNMap}. The region of interest used for the modeling is from $l \in [72.5^\circ, 77.5^\circ]$ and $b \in [-4^\circ, 4^\circ]$ in Galactic coordinates. This field of view includes one of the brightest sources in HAWC's Field of View (FoV), MGRO J2019+368 \citep{hawcj2019}. 

%% FIGURE 3 - CO Dame
\begin{figure*}
    \centering
    \includegraphics[width=0.60\textwidth]{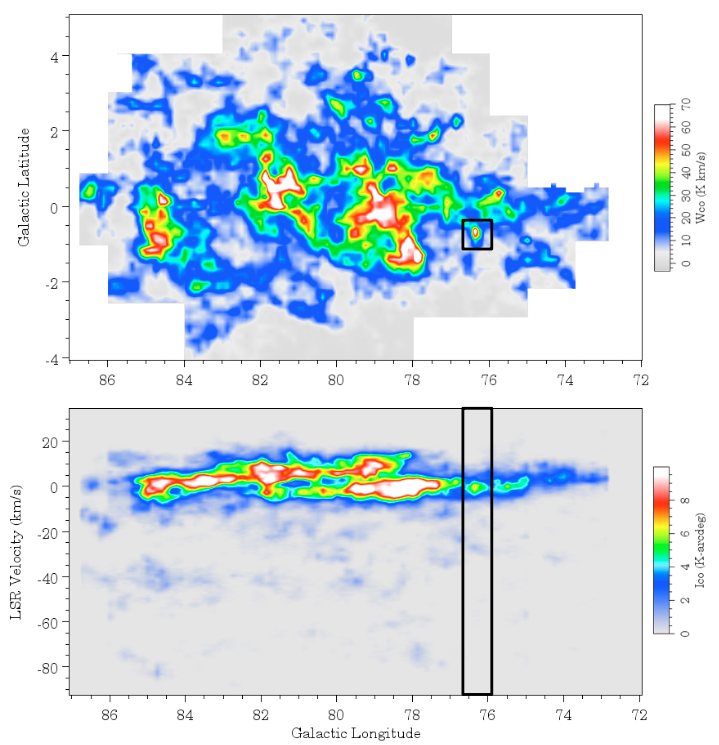}
    \caption{{\footnotesize The 3D spatial distribution of the $^{12}$CO molecular clouds in the ROI from the Dame survey\citep{dame}. The $^{12}$CO region around S106 in black is not surrounded by excess gas in front and behind its location.}}
    \label{fig:dame}   
\end{figure*}

%% Galactic Diffuse Emission Modeling
\subsubsection{Galactic Diffuse Emission Modeling}
\label{sec:GDE_Model}
The Galactic plane is a sophisticated region of diffuse gamma rays, originating from the interaction of CRs with interstellar gas and radiation fields. The diffuse emission coupled with unresolved gamma-ray sources is also known as the Galactic Diffuse Emission (GDE). Since the significance of HAWC J2027 in the NN energy estimator map is $\sim 6.6\sigma$, and the source is at a low galactic latitude, special attention must be paid to the GDE. This analysis employs the use of an energy-dependent GDE template generated with the HERMES software \citep{hermes}, incorporating emission from different radiative processes, Inverse Compton (IC) scattering, Bremsstrahlung, and neutral pion ($\pi^0$) decay from cosmic-ray interactions with atomic (HI) and molecular (H$_2$) hydrogen. The HI and H2 gas maps used as spatial templates are derived from the HI4PI survey \citep{h14pi} and $^{12}$CO Dame Survey \citep{dame}, respectively. For the ($\pi^0$) decay channel, we have adopted the cross-section calculations from \citet{kelneraharonian}, as they are valid for the energy ranges seen by HAWC. A free scaling parameter is applied to the GDE template within the ROI, to allow adjusting its contribution to the local variations in the cosmic-ray density in the overall description of the region.

%% Defining the GDE and S106 Emission in HAWC Data
\subsubsection{Modeling the GDE and S106 Emission in HAWC Data}

The diffuse emission HERMES profile for the region is shown in the left panel of Figure \ref{fig:DiffEmissionMaps}. It is expected that the density of $^{12}$CO (for the molecular hydrogen) and atomic hydrogen is very high at and around the locations of the SFRs. TeV emission observed by HAWC appears as a diffuse morphology source in HAWC's FoV. Modeling the HAWC J2027 source with a gaussian spatial morphology, along with the full radiative emission profile to account for the GDE over the full region, results in a post-trial detection of S106 at $4.8 \sigma$, just below the $5 \sigma$ threshold for source detection in HAWC. From the fit results, we find that this is attributed to the contamination by the bright source MGRO J2019+368 to the overall diffuse emission in the region. 

In order to separate the dense molecular gas emission that correlates with the position of S106 and its vicinity from the diffuse and less dense gas emission in the full ROI, we define a masked emission profile template that masks out a circular region around the location of the dense cloud region near S106 with a radius of 0.5$^\circ$ from the full HERMES diffuse emission profile. This is shown in the left panel of Figure \ref{fig:DiffEmissionMaps} with a red circle. 

Implementing this masked template captures the total gamma-ray emission carrying the fingerprints of the accelerated CRs within this SFR correlating with the molecular clouds in the S106 region.

%% S106 Gas Template Model based on Planck Dust Maps 
\subsubsection{S106 Gas Template Model based on Planck Dust Maps}
\label{sec:Planck}
Assuming that the gamma-ray excess that we observe near S106 is linked to the dense $^{12}$CO clouds near the SFR, we fit a model for J2027 based on the $^{12}$CO observations of the region within the FoV to check and confirm whether the TeV emission observed by HAWC is spatially extended. For the TeV extension model, we use a 2D spatial template based on \href{https://irsa.ipac.caltech.edu/data/Planck/release_2/all-sky-maps/previews/COM_CompMap_Dust-GNILC-Model-Opacity_2048_R2.01/index.html}{GNILC Planck dust opacity map at 353 GHz} shown on the right panel of Figure \ref{fig:DiffEmissionMaps}. The Planck dust opacity map is chosen as it is a more recent observation than the \href{https://lweb.cfa.harvard.edu/rtdc/CO/NumberedRegions/DHT10/index.html}{$^{12}$CO Dame survey}. This template also provides information on HI emission, as well as CO emission, because it shows the optically thick component of the total gas content. The top panel of Figure \ref{fig:dame} shows the 2D spatial distribution of the $^{12}$CO clouds in the vicinity of J2027.

Since the Planck survey does not look into the 3D distribution of dust and clouds, we did a cross check to see if there are any $^{12}$CO cloud distributions either in front or behind the S106 region using the $^{12}$CO Dame survey data. The bottom panel of Figure \ref{fig:dame} shows the spatial velocity distribution of the $^{12}$CO clouds in the direction of S106. From this figure, we see that there is an excess of $^{12}$CO clouds distribution within line of sight velocity range of $-10 < V_{\mathrm{LSR}} / (\rm{km/s}) < 5$, which corresponds to a near and far-kinematic distance of 0.0$-$5.7 kpc calculated using the \href{https://www.treywenger.com/kd/}{tool} described in \citet{kinematicdistancetool}. 

%% FIGURE 4 - HAWC Significance Maps
\begin{figure*}
\centering \vspace*{1pt}
\includegraphics[width=0.48\textwidth]{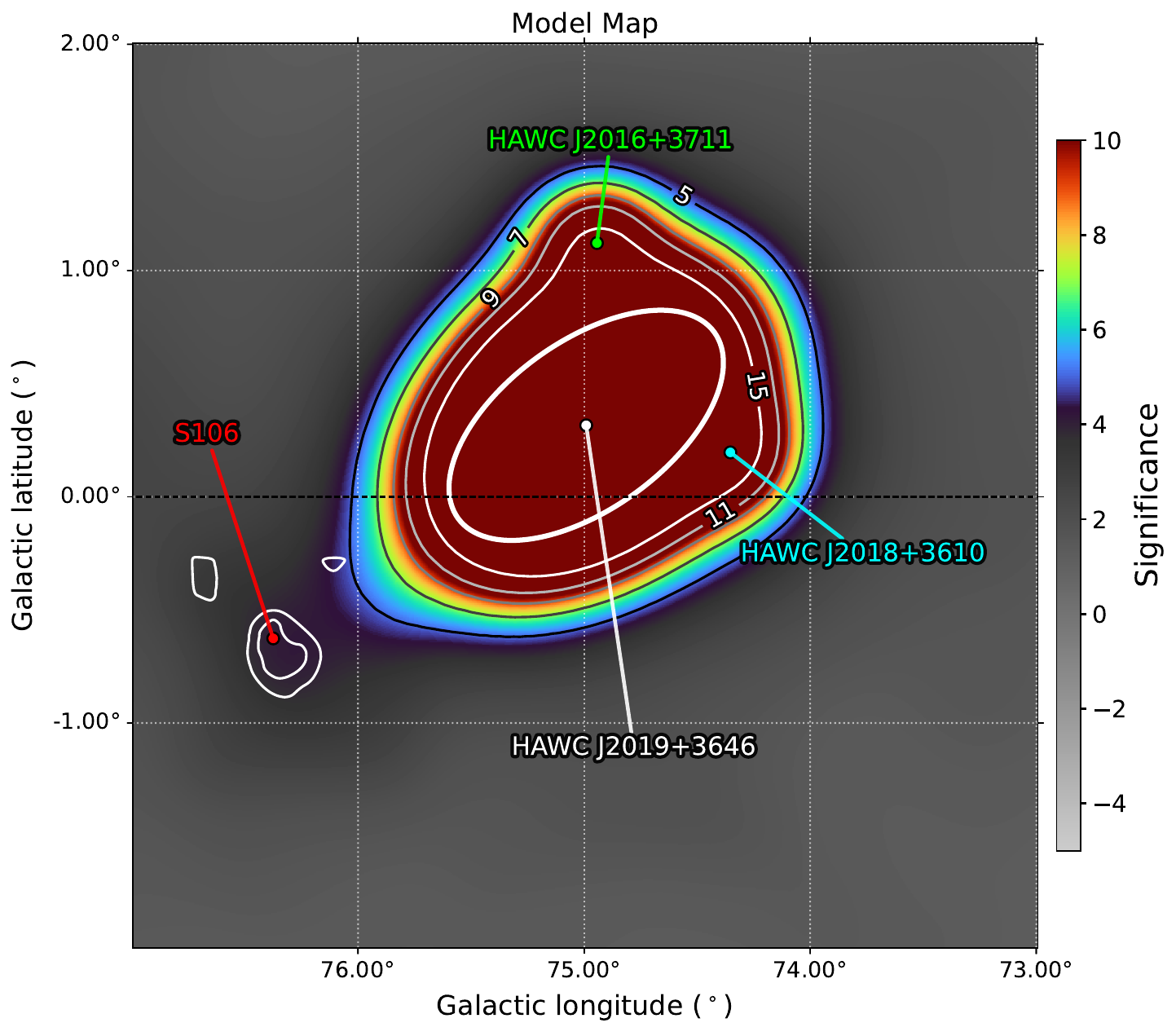}
\includegraphics[width=0.5\textwidth]{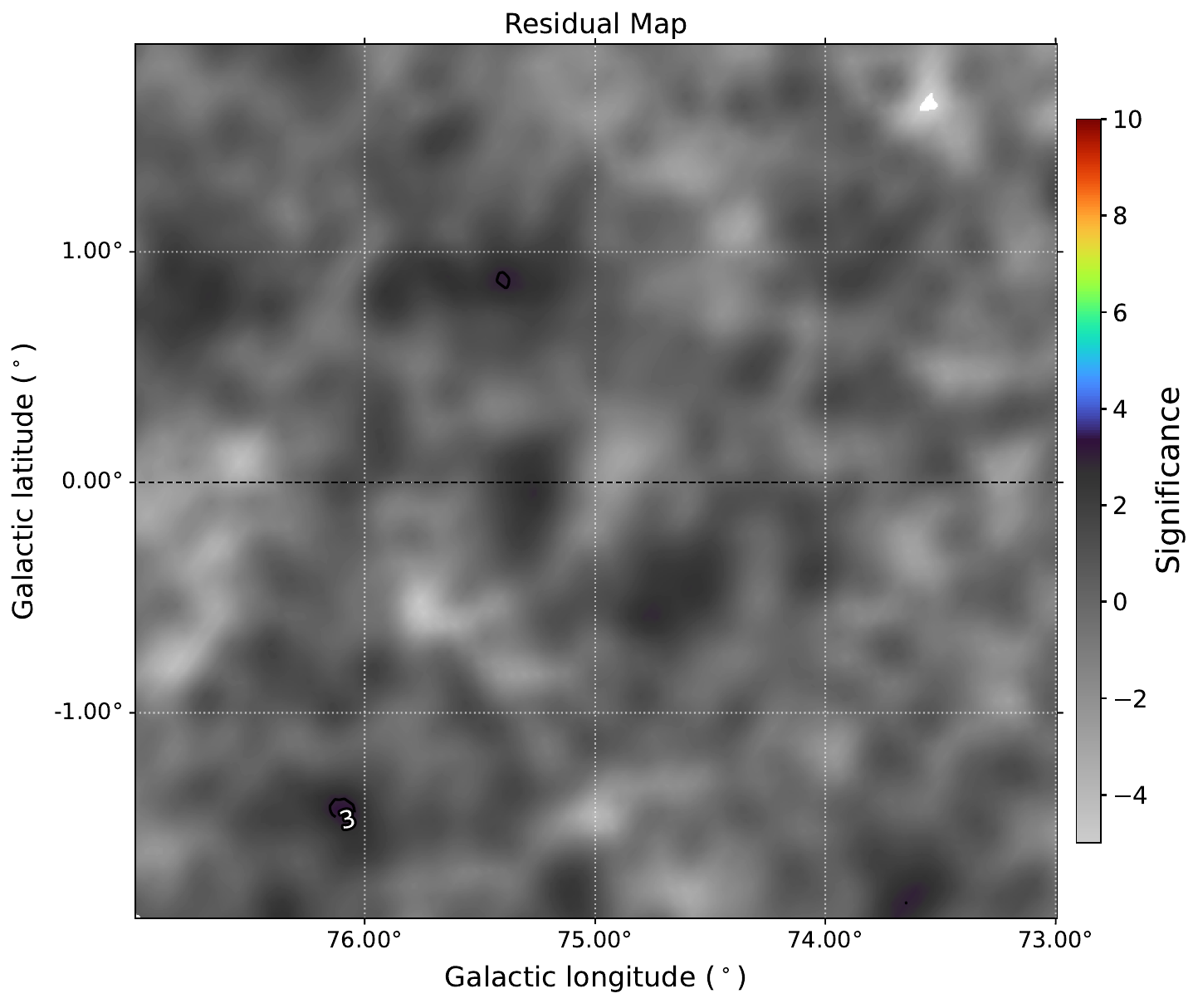}
\includegraphics[width=0.38\textwidth]{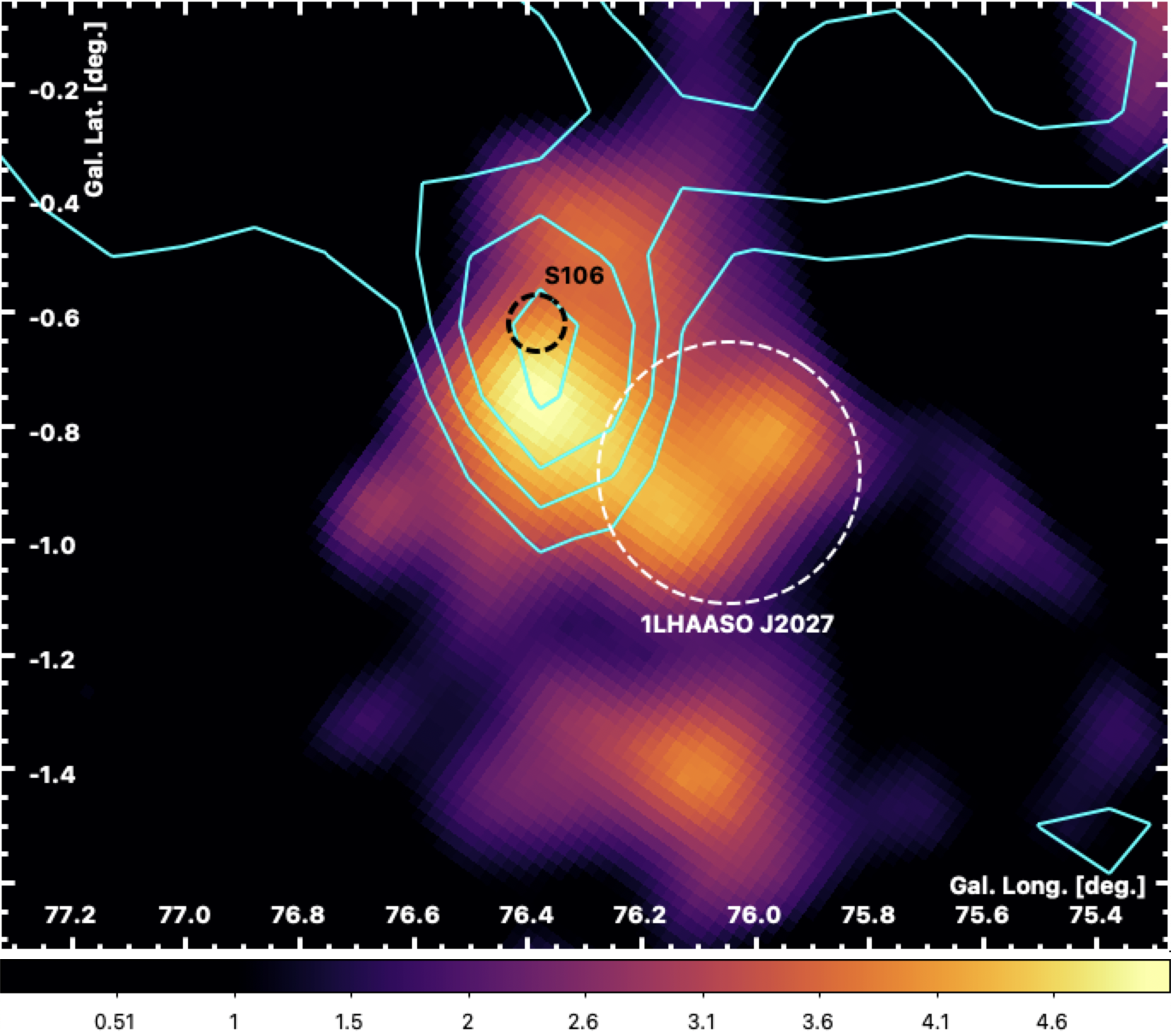}
\includegraphics[width=0.48\textwidth]{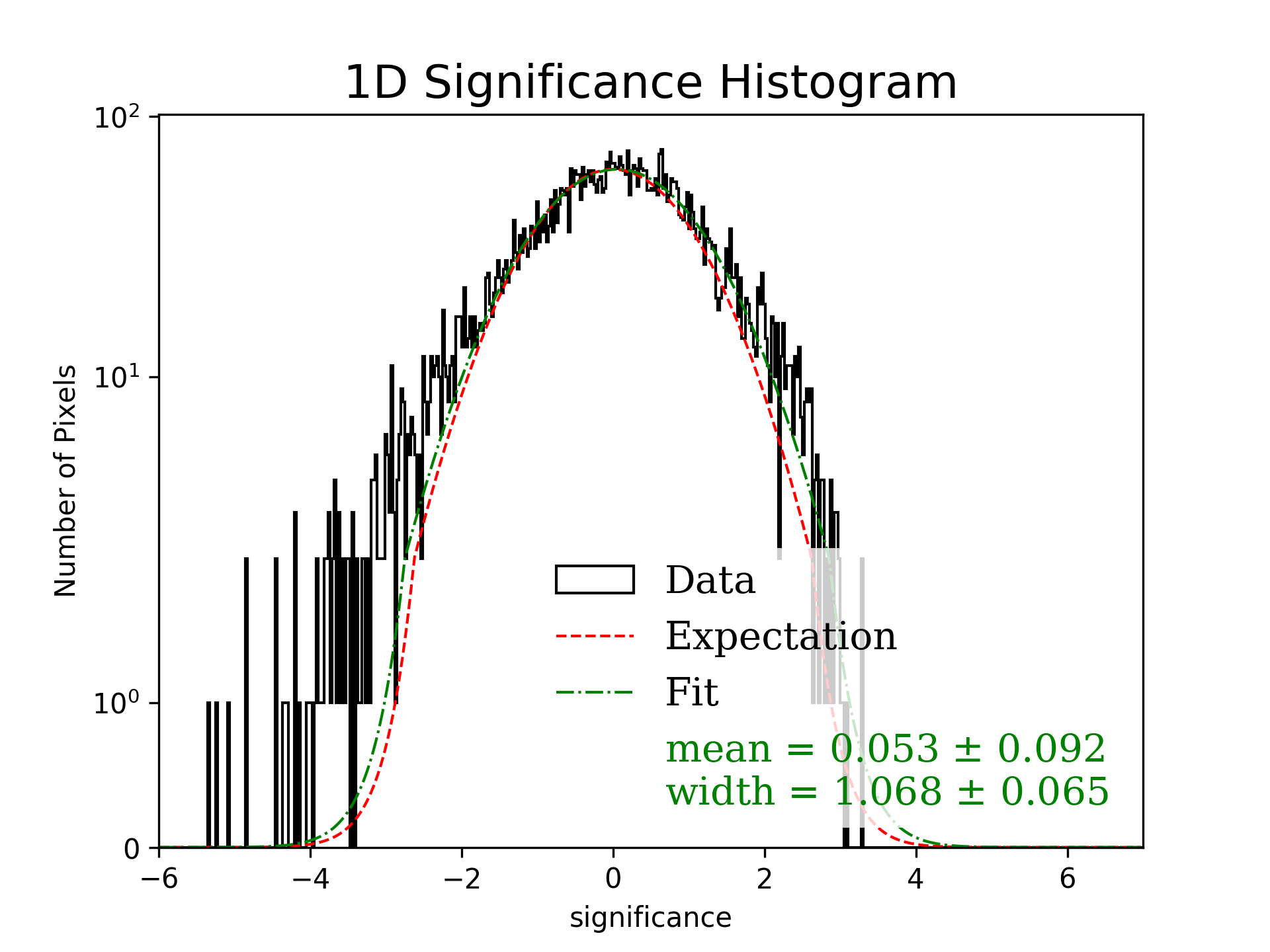}
\caption{ {{\footnotesize {\it Top Left Panel:} The HAWC significance map is showing the location and morphology of the fitted sources in the ROI of J2027. The white contours plotted on HAWC J2027+3718 show the gas morphology represented by the Planck dust template, as described in Section \ref{sec:Planck}. {\it Top Right Panel:} The HAWC residual significance map of the  J2027 ROI after modeling all the sources discussed in Section \ref{sec:HAWCResults}, indicating no excess and no over-subtraction in the modeling. {\it Bottom Left Panel:} The zoomed-in significance map of HAWC J2027+3718, where its color-bar range goes up to 5.1$\sigma$. $^{12}$CO emission contours of 8, 16, 24, 32 K km/s for [-10, 5] km/s taken from the $^{12}$CO Dame survey are shown in cyan colors. In addition, the small black dashed circle shows the position and rough extension of S106 in IR, while the large white dashed circle shows the LHAASO positional error ellipse of J2027. {\it Bottom Right Panel:} The significance histogram (a 1D version of the significance map shown in the top right panel) showing no significant excess remains in the ROI after modeling all sources.} }}
\label{fig:HAWCFitMaps}
\end{figure*}

%%% HAWC Modeling Results
\subsubsection{HAWC Modeling Results}
\label{sec:HAWCResults}
Applying the source-finding algorithm described in \citet{hawcimage} and using the masked HERMES diffuse emission profile template, our best fit model consists of four individual sources: in addition to J2027, we also detect one bright extended source (HAWC J2019+3646 \citep{hawcj2019}) and two point-like sources (HAWC J2016+3711 and the newly discovered HAWC J2018+3610).  Notably, HAWC J2018+3610 is a new unidentified TeV point source.

Initially, we used a point-like source model for J2027, which resulted in the detection of the source above 5$\sigma$ but for all the sources, we also tested alternative spectral and morphological models as described in \citet{hawcic443}. During the morphology test for J2027, the model with the spatial template derived from the Planck dust map (described in Section \ref{sec:Planck}) was preferred over the point-like morphology. We evaluated this preference using the Bayesian Information Criterion \citep{bic} and the Akaike Information Criterion \citep{aic}. The spatial gas template model is favored over the point-like morphology by a $\Delta\mathrm{BIC} = \mathrm{BIC}_{\mathrm{gas}} - \mathrm{BIC}_{\mathrm{point}} = -31$ and $\Delta\mathrm{AIC} = \mathrm{AIC}_{\mathrm{gas}} - \mathrm{AIC}_{\mathrm{point}} =  -6$, where negative values indicate the preference for the spatial gas template because of its lower values. 

The upper left panel of Figure \ref{fig:HAWCFitMaps} shows the sources found in the HAWC analysis and J2027 modeled with the Planck template shown in white contours. Meanwhile, the upper right panel of Figure \ref{fig:HAWCFitMaps} shows the residual map after the subtraction of model from the data map and the lower right panel of Figure \ref{fig:HAWCFitMaps} is the one dimensional binned version of the residual map, which demonstrates that the residual map is devoid of any significant excess, confirming the goodness-of-fit of our model. 

The best-fit parameters of HAWC J2027, the source of interest, are shown in Table \ref{tab:j2027bestfit}, while the best-fit parameters of the other three sources are summarized in Table \ref{tab:roibestfit}. All sources are detected with a significance above our detection threshold of $\sqrt{\mathrm{TS}} > 5\sigma$. 

%% TABLE 1
\begin{table}[ht!]
    \centering
    \begin{tabular}{ccc}
    \hline
        Parameter & Point Source & Planck Template \\
    \hline
    \hline
        RA [deg]& 306.98 $\pm$ 0.04& --  \\
        Dec [deg]& 37.30 $\pm$ 0.03& -- \\
        $\Phi_0$ [(TeV cm$^2$ s)$^{-1}$] & $4.4^{+2.7}_{-1.7}$ & $4.8^{+1.1}_{-0.9}$ $^{+2.3}_{-1.4}$\\
        $\alpha$& 2.499 $\pm$  0.18 & 2.61 $\pm$ 0.11 $^{+0.21}_{-0.19}$\\
        TS & 28.6 & 31  \\
        BIC& 262102.53 & 262072.0  \\
        AIC& 261847.48 & 261841.2  \\
    \hline
    \end{tabular}
    \caption{Best-fit values of HAWC J2027+3718 \\
    NOTE - $\Phi_0$ is the differential flux at E$_0$=7 TeV and is in units of $\times10^{-15}$ [(TeV cm$^2$ s)$^{-1}$]\\
    NOTE - First set of errors is statistical, while the second is systematic. Systematic errors are reported only for the final best-fit model}
    \label{tab:j2027bestfit}
\end{table}

%%% Fermi-LAT Analysis 
\subsection{Fermi-LAT Analysis}
\label{sec:FermiAna}
We re-analyzed archival \textit{Fermi}-LAT data near J2027 to connect the $>\rm{TeV}$ results discussed in Section \ref{sec:HAWCResults} with lower-energy processes. We use \textit{Fermi}-LAT photon data between 4 August 2008 and 20 July 2024 with energies from 200 MeV up to 1 TeV and \texttt{fermitools} version 2.2.0 and \texttt{fermipy} version 1.1.6 \citep{FermiPy}. The ROI for this analysis was circular with radius $20^{\circ}$ centered at J2027's LHAASO position (RA = $306.88^{\circ}$, Dec = $36.95^{\circ}$). The Fermi-LAT instrument response function version P8R3-SOURCE-V3 was used in the analysis while setting the zenith angle maximum at 90$^{\circ}$. 

%% TABLE 2
\begin{table*}[ht!]
    \centering
    \begin{tabular}{lcccc}
    \hline 
        Parameter & J2019+3646 & J2016+3711 & J2018+3610 & GDE\\
    \hline\hline
        RA [deg] & 304.89 $\pm$ 0.01 $^{+0.018}_{-0.017}$ & 304.03 $\pm$ 0.02 $\pm$ 0.043 & 304.57 $\pm$ 0.05 $^{+0.071}_{-0.057}$ & - \\
        Dec [deg] & 36.766 $^{+0.05}_{-0.06}$  $^{+0.01}_{-0.01}$  & 37.19 $\pm$ 0.02 $^{+0.035}_{-0.033}$ & 36.173 $\pm$ 0.03 $^{+0.052}_{-0.046}$ & -\\
        $\sigma$ [deg] & 0.361$\pm$0.01 $^{+0.019}_{-0.014}$& - & - & -\\
        eccentricity & 0.835 $\pm$0.013 $^{+0.024}_{-0.023}$& - & - & -\\
        $\theta$ & 15.6 $\pm$ 1.8 $^{+3.3}_{-2.8}$& - & - & -\\
        $\Phi_0$ (TeV cm$^2$ s)$^{-1}$ & $69^{+9}_{-8}$ $^{+5}_{-6}$ & $10.6^{+2.8}_{-2.2}$ $^{+11.9}_{-3.3}$ & $9.9^{+2.8}_{-2.2}$ $^{+11.9}_{-3.4}$ & $4.02\pm0.24$ $^{+0.46}_{-0.4}$ \\
        $\alpha$ & 1.57$\pm$0.1 $^{+0.431}_{-0.079}$ & 2.61$\pm$0.1 $^{+0.31}_{-0.16}$ & 2.67$\pm$0.11 $^{+0.3}_{-0.18}$ & -\\
        E$_\mathrm{c}$ & $28.2^{+4}_{-3.4}$ $^{+18}_{-3}$ & - & - & -\\
        TS & 2153 & 79 & 41 & 309 \\
    \hline
    \end{tabular}
    \caption{Best-fit values for the three other sources with HAWC Data in the ROI for the Planck Dust Template Model for HAWC J2027+3718. 
    First set of errors is statistical; the second is systematic. 
    $\Phi_0$ is the differential flux at E$_0$ = 2 TeV in units of $\times10^{-14}$ [(TeV cm$^2$ s)$^{-1}$]. 
    $\Phi_0$ for the GDE is the dimensionless scale factor.}
    \label{tab:roibestfit}
\end{table*}

We included all 4FGL-DR4 sources in the ROI in the gamma ray background model, plus the diffuse background model with the Galactic diffuse emission template (\texttt{gll\_iem\_v07.fits}) and the relevant isotropic emission template (iso\_P8R3\_SOURCE\_V3\_v1.txt) obtained at the Fermi science center \citep{LATmodels}. 

With \texttt{gtlike}, we performed the binned maximum likelihood fitting on gamma-ray data. During the fit, we left free the normalization parameters of all the sources within 3$^{\circ}$ ROI with significance $\rm{TS}>20$ as well as the diffuse emission components. The parameters of all the other sources were fixed to values given in the 4FGL-DR4 catalog \citep{4FGLDR4}. 

\begin{figure*}
\centering \vspace*{1pt}
\includegraphics[width=0.99\textwidth]{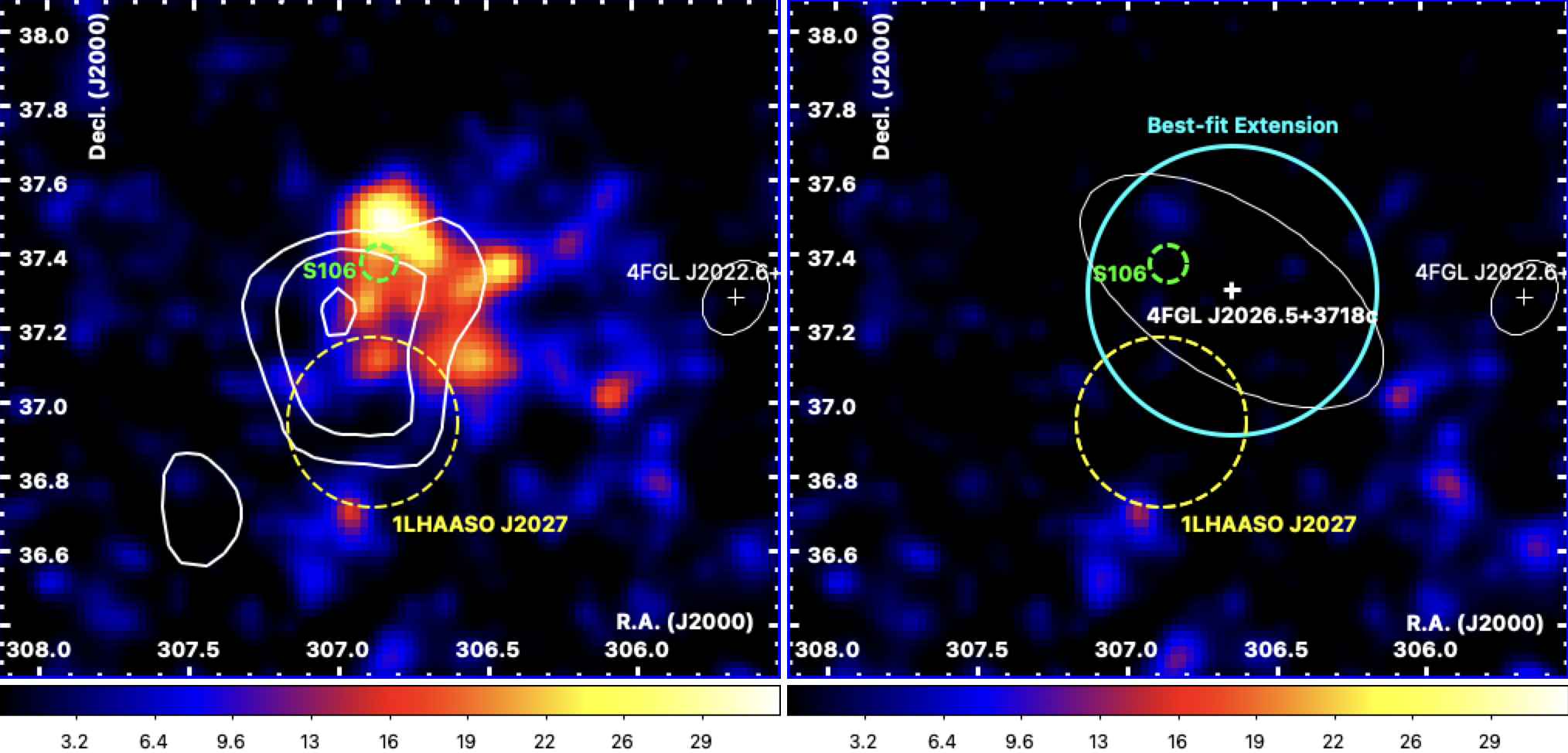}
\caption{{\footnotesize Left Panel: The {\it Fermi}-LAT TS map in the ROI in the energy range of 1 GeV - 1 TeV, where the TS value range is fixed for both maps between 3 and $\sim$32. The position and spatial extent of S106 are indicated by the green dashed circle, while the yellow dashed circle marks the position and positional uncertainty of the LHAASO source. The HAWC significance contours (2.75$\sigma$, 3.5$\sigma$, 4.5$\sigma$) shown in white solid color are taken from the bottom left panel of Figure \ref{fig:HAWCFitMaps}. 
Right Panel: The TS map after adding the best-fit spatial model ({\it RadialGaussian PL}) to the background gamma-ray emission model, after the subtraction of 4FGL J2026.5+3718c (white cross for centroid position, white ellipse as error ellipse). 4FGL J2026.5+3718c with the best-fit extension is represented by the cyan solid circle. }}
\label{fig:FermiSpatModels}
\end{figure*}

%% Fermi-LAT Spatial Analysis 
%\subsubsection{GeV Gamma-ray Morphology}
%\label{sec:FermiSpat}
Figure \ref{fig:FermiSpatModels} shows the zoomed version of the TS map produced for the 10$^{\circ}$ $\times$ 10$^{\circ}$ analysis region in the energy range of 1 GeV $-$ 1 TeV. We computed the best-fit position of 4FGL J2026.5+3718c assuming it to be a point source with log-parabola type spectrum. The best-fit position is found to be $\rm{RA} = 306.82^\circ$, $\rm{Dec} = 37.48^{\circ}$ with a positional uncertainty of $0.04 ^{\circ}$. This new centroid is offset from the 4FGL position by $0.22^{\circ}$.

%% TABLE 3 
\begin{table}
    \caption{Fit results of \textit{Fermi}-LAT data for 4FGL J2026.5+3718c produced in the energy range of 1 GeV $-$ 1 TeV.}
    \begin{minipage}{0.45\textwidth}
    	\centering
	\begin{tabular}{lccc} % creating 3 columns
        \hline\hline
        Spatial Model                     &Spectral Model &$TS_{\rm ext}$  \\
 		\hline
         \\
         \multirow{2}{*}{{\it RadialDisk}}     &Power Law              &38.519 \\
                                               &Broken Power Law             &38.519 \\
                                               &Log-Parabola              &35.684 \\
         \\
         \multirow{2}{*}{{\it RadialGaussian}} &Power Law              &39.456 \\
                                               &Broken Power Law             &39.456 \\
                                               &Log-Parabola              &35.516 \\
        \hline 
        \label{table:FermiExt}
	\end{tabular}
	      \vspace{-0.3cm}
    \end{minipage}
\end{table}

%In addition, it is worth noting that the total TS value of 4FGL J2026.5+3718c between 10 GeV and 1 TeV drops to a value of 5 assuming the source to be a point-source with a log-parabola type spectrum, indicating that gamma rays with energies below 10 GeV are dominating the emission for 4FGL J2026.5+3718c.

% Source Extension Measurements 
To test whether the \textit{Fermi}-LAT emission near J2027 is a point source, we employed the \texttt{extension} module in \texttt{fermipy}, which compares the point-source hypothesis with an extended spatial model using a likelihood-ratio test. For this source, we compared the {\it RadialGaussian} and {\it RadialDisk} spatial models and find a preference for an extended model in all combinations of spatial and spectral models, with $\rm{TS}>25$ in all cases. Table \ref{table:FermiExt} shows a summary of the extension measurements. The best-fit spatial model is {\it RadialGaussian} with a power law (PL) or broken power law (BPL) spectrum, where the Gaussian radius of the extended source is $r = 0.39^{\circ
} \pm 0.07^\circ$ centered around the position of 4FGL J2026.5+3718c.

%% FIGURE 6

%% TABLE 4
\begin{table}
    \caption{The best-fit spatial model for a single extended source (1ES) in the 1 GeV $-$ 1 TeV energy range is found by calculating the $\Delta$AIC value. Column (1) shows the spatial model name. (2)nd and (3)rd columns give the degrees of freedom (d.o.f.) of each model and the corresponding $\Delta$AIC value, respectively. }
    \begin{minipage}{0.45\textwidth}
    	\centering
	\begin{tabular}{lcccc} 
        \hline\hline
        Model   &Spatial Model Name       &d.o.f.          &$\Delta$AIC   \\
        Number  &$~~~~~~$(1)              &(2)             &(3)   \\ 
 		\hline
         \\
         1 &\multirow{1}{*}{\it{1ES: RadialGaussian PL}}   &3         &0   \\
         \\                 
         2 &\multirow{1}{*}{\it{1ES: RadialGaussian BPL}}  &4         &26   
         \\                 
         \hline 
        \label{table:FermiMorph}
        \vspace{-0.3cm}
	\end{tabular}
    \end{minipage}
\end{table}

The resulting TS map after applying the {\it RadialGaussian+PL} model is shown in Figure \ref{fig:FermiSpatModels}. The right panel shows the result of fitting the {\it RadialGaussian+PL} model to the data after the subtraction of 4FGL J2026.5+3718c from the gamma-ray background model.

Using the best-fit spatial extension model described above ({\it RadialGaussian+PL}), we derived the energy spectrum of 4FGL J2026.5+3718c over the energy range from 200 MeV to 1 TeV. We found the total significance of the extended source to be $\sim$11$\sigma$ with the spectral index fitted to $\Gamma$ = 2.18 $\pm$ 0.06. The total flux and energy flux of the extended gamma-ray emission was found to be (1.06 $\pm$ 0.13)$\times$10$^{-8}$ cm$^{-2}$ s$^{-1}$ and (1.09 $\pm$ 0.12)$\times$10$^{-5}$ MeV cm$^{-2}$ s$^{-1}$, respectively. Using the standard \textit{Fermi}-LAT energy bins, we produce the flux points between $200~\:\rm{MeV}$ and $1 \:\rm{TeV}$ shown in Figures \ref{fig:broadSED} and \ref{fig:broadSED2}.

%%% Chandra Analysis 
\subsection{Archival Chandra Data}
\label{sec:Chandra}

There exists a single archival Chandra observation at S106, a $\sim45 \:\rm{ks} $ ACIS-I observation with observation ID 1893. This observation was the subject of a detailed analysis in \cite{2004A&A...424..965G}, studying the X-ray emission from numerous point sources including the central object S106IR. We revisit this archival observation with an eye towards connecting the physical diagnostics enabled with X-ray observations to the VHE emission observed with \textit{Fermi}-LAT and HAWC. While \cite{2004A&A...424..965G} focused on the diverse X-ray point sources in S106, we aimed to evaluate the diffuse X-ray emission near S106 and draw comparisons with the observed high-energy emission. 

\begin{figure*}
    \centering
    \includegraphics[width=0.8\textwidth]{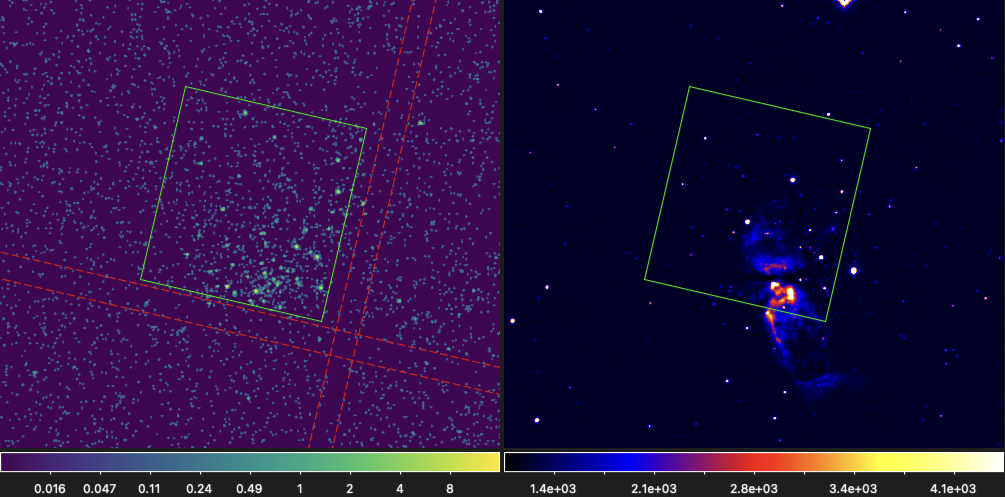}
    \caption{The Chandra ACIS (left) and UKIDSS J-band (right) image of S106 obtained from SkyView, with the source ROI ($2' \times 2' $) in green highlighted in both views. We extract a spectrum by masking out the point sources studied in \cite{2004A&A...424..965G}. The red dashed regions in the Chandra image are the borders of the ACIS chips.}
    \label{fig:XrayIRImage}
\end{figure*}

We downloaded observation 1893 from the Chandra Data Archive, and used the standard \texttt{chandra repro} task in \texttt{ciao} version \texttt{4.17} to reprocess the raw data files into a cleaned event file. Using \texttt{fluximage}, we produced an image of S106 in the ``broad'' ACIS spectral band ($0.5-7.0 \:\rm{keV}$, shown in Figure \ref{fig:XrayIRImage} on the left).

%% FIGURE 7

As we are interested in diffuse, non-thermal emission throughout the S106 complex, we select a $2' \times 2'$ source region enclosing the innermost X-ray point sources in the densest part of S106 (shown in green in Figure \ref{fig:XrayIRImage}) in the north-east ACIS chip and then mask out the known point sources identified in \cite{2004A&A...424..965G} in a five pixel radius. Using a background region of the same size in a source-free region of the same chip, we use \texttt{specextract} to extract a spectrum, response files, and background file for the diffuse emission in S106. We use \texttt{sherpa} version 4.17 to conduct X-ray analysis of this background-subtracted diffuse spectrum.

We adopt an absorbed power law model for the diffuse X-ray flux observed by \textit{Chandra}; the point sources of X-rays are likely young stars \citep{2004A&A...424..965G} and are not related to the relativistic particles studied here. Leaving hydrogen column density $N_H$ a free parameter to account for material close to S106 and in the intervening interstellar space, we obtain a best fit of $N_H = (1.4 \pm 0.5) \times 10^{22} \:\rm{/cm^2}$ and $\Gamma_X = 2.0 \pm 0.4$. The calculated $0.5-8.0 \:\rm{keV}$ flux of the diffuse emission is $(6 \pm 2) \times 10^{-13} \:\rm{erg/s/cm^2}$, or at a distance of $1.4\:\rm{kpc}$ a diffuse X-ray luminosity of $(2.96 \pm 0.5) \times 10^{31} \:\rm{erg/s}$. This is about $10\%$ of the total luminosity from the point sources identified in \cite{2004A&A...424..965G}.

We extract individual flux points from the \textit{Chandra} spectrum by unfolding the spectrum around the measured \textit{Chandra} sensitivity as given in the ARF and RMF files and binning the spectrum to $S/N>5\sigma$ in each bin, shown in Figure \ref{fig:broadSED} and \ref{fig:broadSED2}. We also calculate the synchrotron emission expected with no intervening absorption for comparison.

%%%% Multiwavelength Modeling
\section{Multiwavelength Modeling}
\label{sec:MWLMod}
%%% The Leptonic Model
\subsection{The Leptonic Model}
\label{sec:LeptoMod}

\begin{figure*}
    \centering
    \includegraphics[width=\linewidth]{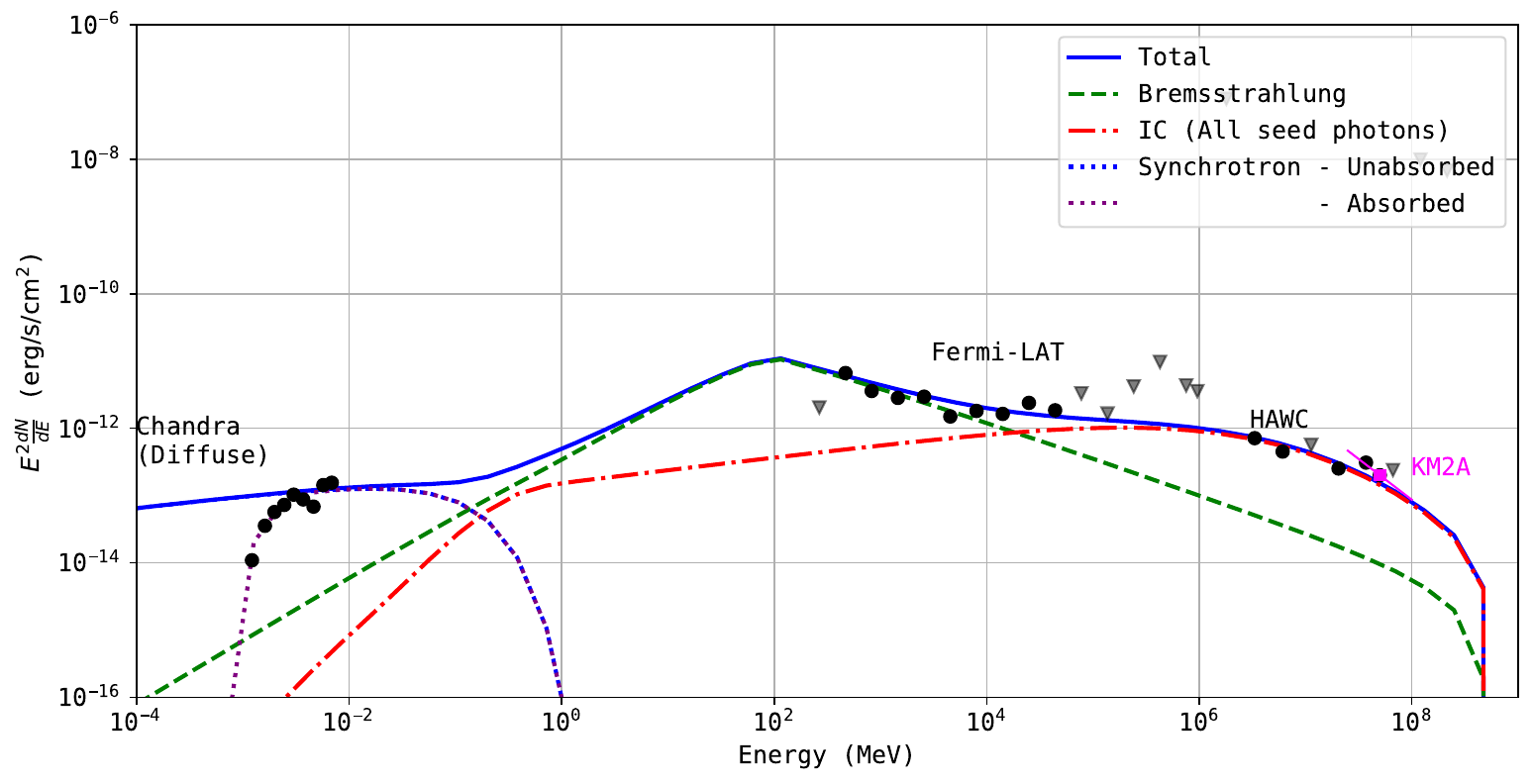}
    \caption{The $E^2 dN/dE$ spectrum of high-energy emission near S106 (black circles for detections, grey arrows for upper limits,  magenta circle for KM2A flux and slope). The fully leptonic model is plotted in blue: synchrotron emission from a $\sim 2.4 \:\rm{\mu G}$ magnetic field (blue and purple dotted), bremsstrahlung (green dashed) in a $n_H = 250 \:\rm{/cm^3}$ cloud, and IC emission (red dot-dash) dominated by IR photons from S106 itself all emerge from a single population of leptons. The sharp cutoff in bremsstrahlung at $\approx 1 \:\rm{GeV}$ is an artifact from minimum particle energy in \texttt{naima}.}
    \label{fig:broadSED}
\end{figure*}

Combining the diffuse X-ray flux from the archival \textit{Chandra} X-ray data with the \textit{Fermi}-LAT and HAWC gamma-ray emission, plus the reported LHAASO-KM2A flux from \cite{2024ApJS..271...25C} and the upper limit implied by non-detection in the most recent \textit{Swift}-BAT point source catalog \citep{2025arXiv250604109L}, we produce the SED in the black points in Figure \ref{fig:broadSED}.

To test whether the \textit{Chandra}, \textit{Fermi}-LAT, and HAWC points could all be simultaneously fit by a single population of high-energy leptons, we develop a model with three discrete emission mechanisms, which we implement using the \texttt{naima} package\citep{naima}. The lepton population is an exponential cutoff power law of the form 

$$ \frac{dN}{dE} = N_0 \left(\frac{E}{\rm{TeV}}\right)^{-\alpha} e^{-\frac{E}{E_{cut}}}$$

These leptons produce emission via the following processes:

\begin{itemize}
    \item $\rm{keV}$ synchrotron radiation produced by leptons cycling around the ambient magnetic field lines in the core of S106, constrained by various previous observations \citep{1988ApJ...335..263K,1995ApJ...442..208R,2005ApJ...627..263V} to be in the range of $10 - 500 \:\rm{\mu G}$. Below $\approx 1\:\rm{keV}$ these X-rays are absorbed by intervening material.
    \item Bremsstrahlung radiation as those same electrons interact with the ambient gas around S106 with a density of $n_H \approx 10^2 - 10^3 /\rm{cm^{3}}$.
    \item IC up-scattering of ambient photons by the same leptons as they stream outwards into the galactic cosmic ray environment.
\end{itemize}

Notably, in addition to the standard photon fields relevant for inverse Compton (IC) emission, the cosmic microwave background (CMB), near- and far-infrared (NIR and FIR) radiation, and local synchrotron photons from synchrotron self-Compton (SSC), the intense infrared radiation field associated with S106 must also be included as a target photon field for up-scattering. We calculate the radiation field density associated with this S106 IR field, by adopting the dust luminosity of \citet{adams2015}, $L_{\rm dust} \gtrsim (9.02 \pm 1.01) \times 10^{4}\,L_{\odot}$, derived from $12-350\:\rm{\mu m}$ photometry of S106 at an assumed distance of $1.4\:\rm{kpc}$ and quoted as a lower limit because some stellar radiation may escape through the bipolar cavity. The radiation energy density at a distance $r$ from S106 is $u_{\rm rad} = L_{\rm dust}/(4\pi r^{2} c)$, which gives $3.06 \times 10^{-5}\:\rm{eV\,cm^{-3}}$ at Earth and $\approx 60\:\rm{eV\,cm^{-3}}$ at $r = 1\:\rm{pc}$. We evaluate this expression at the characteristic distance $d_{IC}$ from S106 at which the IC up-scattering by the lepton population occurs, treating $d_{IC}$ as a free parameter in the fit.

All the photon fields used in the IC model are described in Table \ref{tab:ICfields}, including their blackbody temperature $T_{\rm{bb}}$ and energy density $u_{\rm rad}$. The S106 field exceeds the combined CMB, FIR and NIR energy density ($1.76\:\rm{eV\,cm^{-3}}$) inside $\approx 5.8 \:\rm{pc}$, and falls below it at larger separations. Additionally, if the observed HAWC and LHAASO emission is produced by IC up-scattering on these spatial scales, the extension seen in $\rm{TeV}$ gamma-rays can be naturally reproduced.

% \begin{deluxetable}{lccc}% C for mathmode
% \tablecaption{Fields invoked for IC upscattering. $d_{IC}$ is the characteristic distance from S106 where IC upscattering with takes place.} \label{tab:ICfields}
% \tablewidth{\columnwidth}
% \tablehead{
% \colhead{} & \colhead{Field} & \colhead{$T_{bb}$} & \colhead{$\rho_{e}$} \\
% \colhead{} & \colhead{} & \colhead{$(\rm{K})$} & \colhead{$(\rm{eV/cm^3})$} }
% \startdata
% & CMB & 2.7 & 0.261 \\
% & FIR & 30 & 0.5 \\
% & S106 & 100 & $0.04 \: (d_{IC}/\rm{kpc})^{-2} $ \\
% & NIR & 3000 & 1.0 \\
% & SSC & & (X-ray synchrotron field) \\
% \enddata
% \end{deluxetable}
\begin{deluxetable}{lcc}
% \color{red}
\tablecaption{Photon fields invoked for IC upscattering. $d_{IC}$ is the
characteristic distance from S106 at which the upscattering {\bf of IR photons} takes place. {\bf SSC has a negligible energy density relative to the other photon fields.}}
\label{tab:ICfields}
\tablewidth{\columnwidth}
\tablehead{
\colhead{Field} & \colhead{$T_{bb}$} & \colhead{$u_{\rm rad}$} \\
\colhead{} & \colhead{$(\rm{K})$} & \colhead{$(\rm{eV\,cm^{-3}})$} }
\startdata
CMB  & 2.7  & 0.261 \\
FIR  & 30   & 0.5 \\
S106 & 100  & $60 \: (d_{IC}/\rm{pc})^{-2}$ \\
NIR  & 3000 & 1.0 \\
SSC  & \nodata & (X-ray synchrotron field) \\
\enddata
\tablecomments{ The S106 field is normalized from the dust luminosity
$L_{\rm dust} \gtrsim (9.02 \pm 1.01) \times 10^{4}\,L_{\odot}$ measured by
\citet{adams2015} at $d = 1.4$~kpc, via
$u_{\rm rad} = L_{\rm dust}/(4\pi r^{2} c)$.} 
\end{deluxetable}

It is worth noting that each of the emission mechanisms noted above have a single astrophysical parameter that can scale up or down the flux from that particular mechanism ($B$ for synchrotron, $n_H$ for bremsstrahlung, and the additional IR field from S106 for IC). Astrophysical modeling can constrain each of these parameters, but the extended emission could also occur in diverse environments around S106 with varying values for each.

In our modeling, we fix the ambient gas density at $n_H = 250 \:\rm{/cm^3}$ and allow the magnetic field $B$, the characteristic IC scattering distance $d_{IC}$, and the index, cutoff energy and normalization of the lepton distribution to vary. The model parameters are fitted to the multi-wavelength data using the Markov Chain Monte Carlo (MCMC) implementation within \texttt{naima}'s framework, which samples the parent particle population parameters. We note that the parameters of theses models are correlated and are/or remain poorly constrained by the available data. Consequently, the best-fit parameter values should not be interpreted as independent measurements of the physical quantities, but rather as a plausible set of values that provides a reasonable description of the observed data. The best fit parameter values are summarized in Table \ref{tab:Lparam} and the resulting spectrum is shown in Figure \ref{fig:broadSED}. The total energy in the electron population above $1\:\rm{GeV}$, $W_e$, is derived from the best-fit values of $N_0$, $E_{\rm cut}$, and $\alpha$.
The best-fit $d_{IC} = 3.55\:\rm{pc}$ therefore places the up-scattering region where photons from S106 itself supply the dominant seed population, exceeding the ambient fields by a factor of $\approx 2.7$. Notably, this crossover radius is comparable to the observed extent of the TeV emission ($0.25^{\circ}$, or $\approx 6.1\:\rm{pc}$ at $1.4\:\rm{kpc}$), so the size of the IC-emitting region is set by the scale over which S106's own radiation field dominates.

% \begin{deluxetable}{lc|l}% C for mathmode
% \tablecaption{Parameters for the leptonic model in Figure \ref{fig:broadSED}.} \label{tab:Lparam}
% \tablewidth{\columnwidth}
% \tablehead{ \colhead{Parameter} & \colhead{units} & \colhead{value} }
% \startdata
% $N_0$ & $\rm{1/eV}$ & $2.96 \times 10^{31}$  \\
% $E_{\rm{cut}}$ & $\rm{TeV}$ & 1000  \\
% $\alpha$ & & 2.52  \\
% $W_e$ & $\rm{erg}$ & $3.5 \times 10^{45}$ & & \\
% $B$ & $\rm{\mu G}$ & 4.87 \\
% $n_H$ & $\rm{1/cm^3}$ & 250  \\
% $d_{IC}$ & $\rm{pc}$ & 30 \\
% \enddata
% \end{deluxetable}

% \begin{deluxetable}{lc|l}% C for mathmode
% \tablecaption{Parameters for the leptonic model in Figure \ref{fig:broadSED}.} \label{tab:Lparam}
% \tablewidth{\columnwidth}
% \tablehead{ \colhead{Parameter} & \colhead{units} & \colhead{value} }
% \startdata
% $N_0$ & $\rm{1/eV}$ & $6.16^{+4.89}_{-1.34} \times 10^{31}$  \\
% $E_{\rm{cut}}$ & $\rm{TeV}$ & $1753^{+3005}_{-1179}$  \\
% $\alpha$ & & $2.412^{+0.042}_{-0.095}$  \\
% $W_e$ & $\rm{erg}$ & $4.11^{+0.50}_{-0.40} \times 10^{45}$  \\
% $B$ & $\rm{\mu G}$ & $2.38^{+0.65}_{-0.99}$ \\
% $n_H$ & $\rm{1/cm^3}$ & 250 \\
% $d_{IC}$ & $\rm{pc}$ & $3.55^{+4.99}_{-3.42}$ \\
% \enddata
% \end{deluxetable}

\begin{deluxetable}{lc|l}% C for mathmode
% \color{red}
\tablecaption{Parameters for the leptonic model in Figure \ref{fig:broadSED}.} \label{tab:Lparam}
\tablewidth{\columnwidth}
\tablehead{ \colhead{Parameter} & \colhead{units} & \colhead{value} }
\startdata 
$N_0$ & $\rm{1/eV}$ & $6.16 \times 10^{31}$  \\
$E_{\rm{cut}}$ & $\rm{TeV}$ & $1753$  \\
$\alpha$ & & $2.41$  \\
$W_e$ & $\rm{erg}$ & $4.11 \times 10^{45}$ \\
$B$ & $\rm{\mu G}$ & $2.38$ \\
$n_H$ & $\rm{1/cm^3}$ & $250$ \\
$d_{IC}$ & $\rm{pc}$ & $3.55$ \\
\enddata
\end{deluxetable}

%% FIGURE 8

As shown in Figure \ref{fig:broadSED}, a single-population leptonic model can fit the entire SED for S106, from the \textit{Chandra} diffuse emission to the HAWC and KM2A data points. If a population of high-energy hadrons is present, it would be sub-dominant at all relevant energies.

%%% The Lepto-hadronic Model
\subsection{The Lepto-hadronic Model}
\label{sec:LeptHadroMod}

\begin{figure*}
    \centering
    \includegraphics[width=\linewidth]{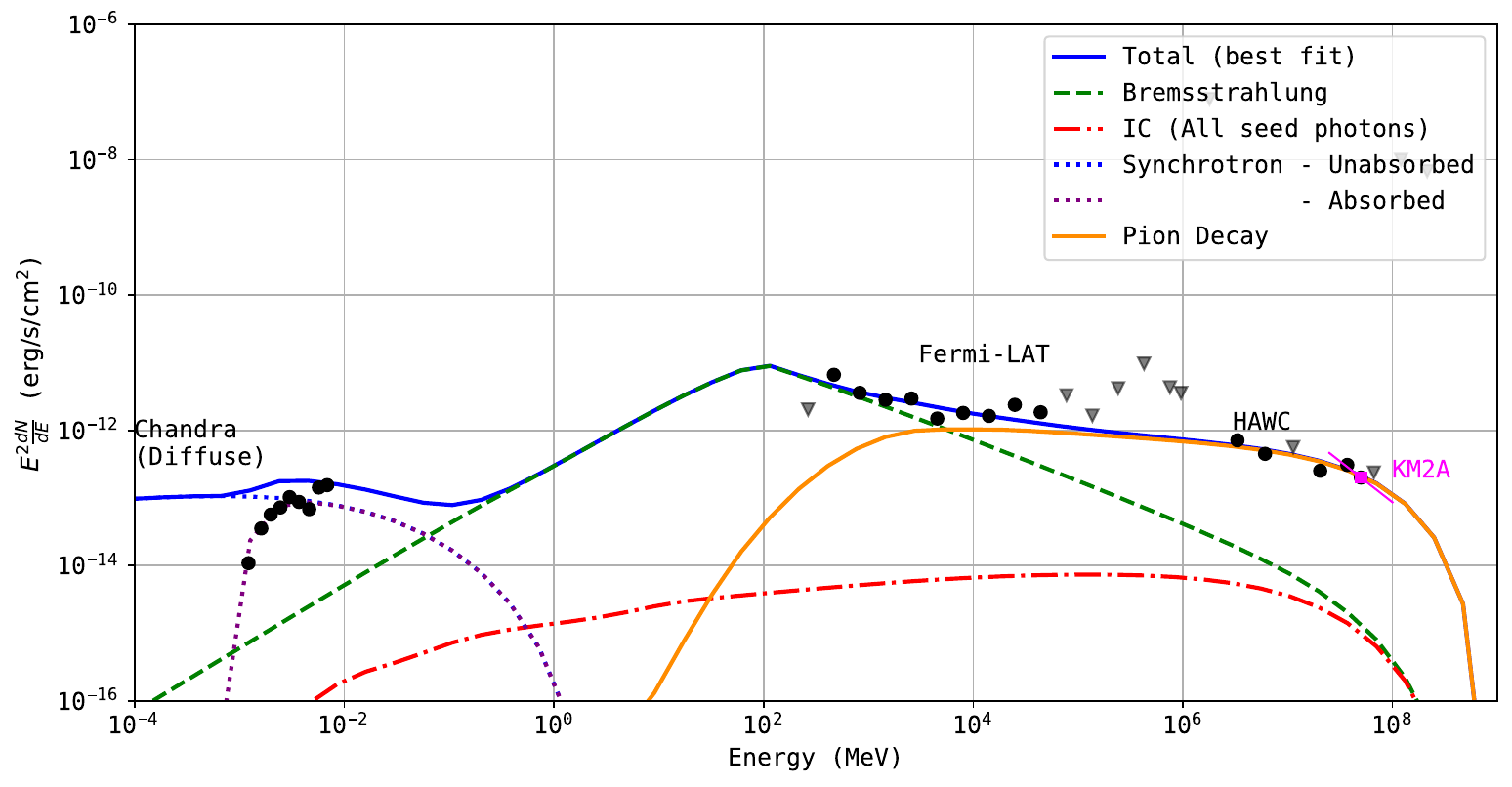}
    \caption{The $E^2 dN/dE$ spectrum of high-energy emission near S106, but with a lepto-hadronic model. Instead of IC emission, pion decay (solid orange) produces the $>\rm{TeV}$ gamma-ray flux. In this case, the IC emission (red dash-dot) is subdominant at $\rm{TeV}$ energies.}
    \label{fig:broadSED2}
\end{figure*}

In an alternative hybrid lepto-hadronic model, X-rays are still produced via diffuse synchrotron emission in S106, and $\rm{GeV}$ gamma-rays are still the product of lepton bremsstrahlung, but the gamma rays are produced by the decay of hadrons into gamma-rays via neutral pions. Figure \ref{fig:broadSED2} shows one possible lepto-hadronic model. The addition of a hadronic component introduces many new free parameters into the model, such as the energy density, slope, and cutoff of the hadron distribution while more weakly constraining the lepton distribution.

Figure \ref{fig:broadSED2} shows the broadband SED in this lepto-hadronic model, with gamma rays from pion decay explaining the $\rm{TeV}$ emission instead of IC scattering. This is achieved by increasing the effective distance at which the IC emission occurs with respect to S106, thereby decreasing the energy density of the IR field from S106 and greatly reducing the IC luminosity of the system. The $>\rm{TeV}$ gamma-rays are then entirely the product of hadronic pion decay, with IC emission being subdominant at all relevant energies.

\begin{deluxetable}{lc|l}% C for mathmode
% \color{red}
\tablecaption{Parameters for the lepto-hadronic model in Figure \ref{fig:broadSED2}.} \label{tab:LHparam}
\tablewidth{\columnwidth}
\tablehead{ \colhead{Parameter} & \colhead{units} & \colhead{value} }
\startdata
 Leptons & & \\
$N_0$ & $\rm{1/eV}$ & $1.878 \times 10^{31}$  \\
$E_{\rm{cut}}$ & $\rm{TeV}$ & 99.5  \\
$\alpha$ & & 2.7 \\
$W_e$ & $\rm{erg}$ & $5.3 \times 10^{45}$\\
 & & \\
 \hline
 Hadrons & & \\
$N_0$ & $\rm{1/eV}$ & $1.8 \times 10^{33}$  \\
$E_{\rm{cut}}$ & $\rm{TeV}$ & 1000  \\
$\alpha$ & & 2.16  \\
$W_p$ & $\rm{erg}$ & $3.1 \times 10^{46}$ \\
 & & \\
$B$ & $\rm{\mu G}$ & 18.2 \\
$n_H$ & $\rm{1/cm^3}$ & 250  \\
$d_{IC}$ & $\rm{pc}$ & 15.1 \\
\enddata
\end{deluxetable}

Table \ref{tab:LHparam} gives the parameters of the lepto-hadronic model, once again calculating the energies in the leptons and hadrons $W_e$ and $W_p$ above $1\:\rm{GeV}$. The hadrons in this model contain substantially more energy than the leptons, by dint of their higher cutoff energy and shallower spectrum.

%% FIGURE 9

%%%% Discussion and Conclusions
\section{Discussion and Conclusions}
\label{sec:DiscConc}
%%% Discussion
\subsection{Discussion}
\label{sec:Disc}

Our fully leptonic model for the combined \textit{Chandra}, \textit{Fermi}-LAT, HAWC, and LHAASO data points in Figure \ref{fig:broadSED} has a boosted ambient IR field close to S106 contributing most seed photons to $>\rm{TeV}$ IC emission. Considering this model, it would seem that the gamma-ray emission from hadronic components is not strictly needed. This is not to say that emission above $\approx\rm{TeV}$ cannot have hadronic components; In the lepto-hadronic model fit shown in Figure \ref{fig:broadSED2}, lowering the leptonic contributions to the $\:\rm{TeV}$ flux while increasing the hadronic contribution results in a hybrid model. 

The critical factors differentiating between the leptonic and hadronic models are the magnitude of the ambient photon fields for IC up-scattering and the total energy in the hadron population. If the ambient photon fields for IC up-scattering are high, perhaps because the scattering occurs very close the S106's bright IR emission, then the gamma rays will be almost entirely leptonic and take place over an extended region. If the $\rm{TeV}$ gamma rays are hadronic in origin they would be correlated with the density of target matter throughout the region near S106.  

%% Cooling Timescales
\subsubsection{Cooling Timescales}

Is it feasible that a single population of leptons dominates the extended X-ray, GeV, and TeV emission observed near S106? The cooling timescales of each of these processes must be larger than the physical size of S106, such that a single population of leptons can cause all three emission mechanisms without depleting its reservoir of energy. S106 has a size on the sky of $\approx 2'$, which at a distance of $\approx 1.4 \:\rm{kpc}$ is a physical size of $\approx 3 \:\rm{ly}$.  The cooling timescales of the emission mechanisms in our leptonic model must all be substantially greater than the light crossing time of S106 to ensure that a lepton population can participate in all three mechanisms.

For synchrotron emission, we compare the energy of a typical electron $E = \gamma_0 m_e c^2$ with the synchrotron power radiated by that electron $P_{syn} = 2 \gamma_0^2\sigma_T c U_B \rm{sin^2(\alpha)}$. Adopting the pitch angle average $\langle\rm{sin^2(\alpha)}\rangle = 2/3$, together with $B = 2.38 \:\rm{\mu G}$ and  conservatively, a Lorentz factor for the electrons of $\gamma = 10^9$ (corresponding to a lepton energy $E \approx 0.5 \:\rm{PeV}$), we obtain the characteristic synchrotron cooling timescale of, 

$$ t_{syn} = \frac{E}{P_{syn}} = \frac{m_e c^2}{\frac{4}{3}\sigma_Tc\gamma U_B} \approx 4329 \:\rm{yr}$$

where $m_e$ is the mass of the electron, c is the speed of light, $\sigma_T$ is the Thomson cross section, and $U_B$ is the magnetic field energy density given by $U_B = B^2 / 8 \pi$.

For bremsstrahlung, \cite{2014PhRvD..90b3526A} and \cite{2021MNRAS.504.1840C} give the cooling time as, with our assumed $n_H = 250/\rm{cm^3}$,

$$t_{brem} = 4 \times 10^{7} \:\rm{yr} ~ \left(\frac{n_H}{\rm{cm}^3} \right)^{-1} = 160000 \:\rm{yr} $$

The cooling timescale for IC scattering in the Klein--Nishina (KN) regime, $t_{\rm IC}$, is given by \citep{moderski}:

$$
t_{\rm IC}(\gamma) =
\frac{3m_e c^2}
{4c\sigma_T \gamma
\displaystyle\sum_i
\left[ \frac{U_{{\rm rad},i}} {\left(1+4\gamma\varepsilon_{0,i}\right)^{3/2}}\right]
}
$$

Here, $U_{{\rm rad},i}$ and $\varepsilon_{0,i}$ are the energy density of the $i$th radiation field and the mean energy of its seed photons, respectively, where $\varepsilon_{0,i}$ is expressed in units of $m_ec^2$. Using the energy densities and mean photon energies listed in Table~\ref{tab:ICfields}, with the latter calculated from the Planck blackbody relation $\langle E\rangle = 2.7k_{\rm B}T$, we obtain an IC cooling timescale of $t_{\rm IC}=28.6~\rm{kyr}$. We note that at $\gamma = 10^9$, the \textbf{IR photons from S106} and NIR fields are strongly Klein--Nishina suppressed ($4\gamma\varepsilon_0 \gg 1$), so the IC cooling of the highest-energy electrons is dominated by the CMB even though the S106 field dominates the total radiation energy density at $d_{IC}$.

All three of these timescales are large enough such that a single population of  leptons streaming out at relativistic speeds will survive to produce the three emission mechanisms in the leptonic model before escaping into the wider galaxy as cosmic rays.

%% Total Energy Budget
\subsubsection{Total Energy Budget}

Young stellar clusters are often invoked as sources of cosmic rays, the sites of dramatic particle acceleration. We have shown that it is feasible for X-ray, GeV, and TeV emission from systems like S106 to be attributed to fully leptonic processes. For the leptonic model we can also compare the total lepton energy budget $W_e = 4.1 \times 10^{45}\:\rm{erg}$ to the kinetic energy of the outflows from the central source of S106, thought to be a young O+B binary named S106IR \citep{Comeron_2018}.

The total kinetic energy luminosity of the outflows from S106IR is given by

$$ L_k = \frac{1}{2} \dot M v_{\rm{flow}}^2 $$

\noindent where $\dot M$ is the mass loss rate from S106IR and $v_{\rm{flow}}$ is the measured velocity of the winds. \cite{1982BAAS...14..925S} and \cite{1983ApJ...272..154B} find $\dot M \approx 10^{-6} M_\odot/\rm{yr}$ and $v_{\rm{flow}} \approx 150 \:\rm{km/s}$, though admittedly both works arrived at these results before the likely binary nature of S106IR was established. Using those values, we calculate $L_k \approx 2 \times 10^{41} \:\rm{erg/year}$ in the outflows from S106IR.

For the duration over which this luminosity is sustained, several estimates can be applied. An upper limit is the age of the S106 complex as a whole, estimated in \cite{2006AJ....131.1608O} as roughly $\sim \rm{Myr}$. The lower limit might be the possible supersonic explosion $3500$ years ago cited in \cite{Bally_2022}. Adopting the rather conservative IC cooling time noted above $t_{IC} = 28.6\:\rm{kyr}$ we obtain a total reservoir of kinetic energy over that period of $E_k = t_{IC} \times L_k \approx 5.72 \times 10^{45}\:\rm{erg}$. The order-of-magnitude similarity of this estimate for a single high-mass system to the energy stored in the lepton distribution supports the at-large shocked outflows from multiple young stars in S106 as the engine behind the high-energy fluxes analyzed here.

%% Particle Diffusion Coefficients
\subsubsection{Particle Diffusion Coefficients}

To evaluate whether particle transport through the S106 region is feasible,
we parameterize the standard energy-dependent diffusion coefficient as
\[
D(E)=\chi D_{0}\left(\frac{E}{E_{0}}\right)^{\delta},
\]
where \(\chi\) accounts for suppression relative to the average Galactic diffusion coefficient, and \(\delta\) describes the turbulence regime \citep{Ptuskin2006,Blasi2013}. We estimate the effective transport coefficient required for particles to reach an offset emission region by equating the observed propagation scale \(R\) with the diffusion radius,
\[
R_{\rm diff}(E,t)=\sqrt{4D(E)t}.
\]
This gives
\[
D_{\rm req}(R,t)\sim \frac{R^{2}}{4t},
\]
following the standard Green's-function solution of the diffusion equation for particles escaping from a compact accelerator \citep{Atoyan1995,Ohira2011}. Therefore,
\[
D_{\rm req}
\simeq
7.6\times10^{27}
\left(\frac{R}{100~{\rm pc}}\right)^2
\left(\frac{t}{10^{5}~{\rm yr}}\right)^{-1}
{\rm cm^{2}~s^{-1}} .
\]
Here \(D_{\rm req}\) is not an independently fitted quantity; rather, it is the effective diffusion coefficient required for particles to reach the observed distance \(R\) within an activity time \(t\). We compare this value with the Bohm diffusion coefficient,
\[
D_{\rm B}(E)=\frac{1}{3}r_{\rm L}c
\simeq
3.3\times10^{25}
\left(\frac{E}{1~{\rm TeV}}\right)
\left(\frac{1~\mu{\rm G}}{B}\right)
{\rm cm^{2}~s^{-1}},
\]
where \(r_{\rm L}\simeq E/(qB)\) is the Larmor radius for a relativistic particle. Bohm diffusion corresponds to the limiting case in which the particle mean free path is comparable to \(r_{\rm L}\), and therefore \(D_{\rm B}(E)\propto E\), i.e. \(\delta=1\).

For the purely leptonic fit, the best-fit IC distance is
% \[
% d_{\rm IC}=3.55^{+4.99}_{-3.42}~{\rm pc},
% \]
\[
d_{\rm IC}=3.55~{\rm pc},
\]
with
\[
B=2.38~\mu{\rm G}, \qquad
E_{{\rm cut},e}=1753~{\rm TeV},
\]
and
\[
W_{e}=4.11\times10^{45}~{\rm erg}.
\]
For this compact best-fit distance, the required diffusion coefficient is
\[
D_{\rm req}^{\rm lep}
\simeq
9.6\times10^{24}
\left(\frac{d_{\rm IC}}{3.55~{\rm pc}}\right)^2
\left(\frac{t}{10^{5}~{\rm yr}}\right)^{-1}
{\rm cm^{2}~s^{-1}} .
\]
At the fitted electron cutoff energy, the Bohm coefficient is
\[
D_{{\rm B},e}^{\rm lep}(1753~{\rm TeV})
\simeq
2.4\times10^{28}
\left(\frac{2.38~\mu{\rm G}}{B}\right)
{\rm cm^{2}~s^{-1}} .
\]
The synchrotron cooling time is approximately
\[
t_{\rm syn}
\simeq
1.3\times10^{7}
\left(\frac{B}{1~\mu{\rm G}}\right)^{-2}
\left(\frac{E_{e}}{1~{\rm TeV}}\right)^{-1}
{\rm yr}.
\]
For \(B=2.38~\mu{\rm G}\) and \(E_{e}=1753~{\rm TeV}\), this gives \(t_{\rm syn}\sim1.3\times10^{3}~{\rm yr}\). The corresponding Bohm diffusion length is of order \(\sim20~{\rm pc}\), larger than the best-fit \(d_{\rm IC}\). Therefore, the updated purely leptonic solution is transport-compatible in the sense that the IC-emitting region is inferred to lie only a few parsecs from S106. This favors an interpretation in which the Fermi-LAT Bremsstrahlung component traces lower-energy electrons confined in the dense S106 environment, while the IC component is produced relatively close to the acceleration site.

We next consider the lepto-hadronic interpretation. In this model, the electron and proton spectra are described by power laws with exponential cutoffs, with
\[
\alpha_{e}=2.70, \qquad
E_{{\rm cut},e}=99.5~{\rm TeV},
\]
and
\[
\alpha_{p}=2.16, \qquad
E_{{\rm cut},p}=1000~{\rm TeV}.
\]
The corresponding total particle energies are
\[
W_{e}=5.300\times10^{45}~{\rm erg}, \qquad
W_{p}=3.081\times10^{46}~{\rm erg},
\]
and the fitted magnetic field is
\[
B=18.22~\mu{\rm G}.
\]
As a compact lepto-hadronic transport estimate, we first evaluate the required diffusion coefficient for
\[
d_{\rm IC}=3.55~{\rm pc},
\]
matching the compact scale favored by the purely leptonic fit. This gives
\[
D_{\rm req}^{\rm LH,3.55}
\simeq
9.6\times10^{24}
\left(\frac{d_{\rm IC}}{3.55~{\rm pc}}\right)^2
\left(\frac{t}{10^{5}~{\rm yr}}\right)^{-1}
{\rm cm^{2}~s^{-1}} .
\]
For the lepto-hadronic magnetic field, the Bohm coefficients at the fitted cutoff energies are
\[
D_{{\rm B},e}^{\rm LH}(99.5~{\rm TeV})
\simeq
1.8\times10^{26}
\left(\frac{18.22~\mu{\rm G}}{B}\right)
{\rm cm^{2}~s^{-1}},
\]
and
\[
D_{{\rm B},p}^{\rm LH}(1000~{\rm TeV})
\simeq
1.8\times10^{27}
\left(\frac{18.22~\mu{\rm G}}{B}\right)
{\rm cm^{2}~s^{-1}} .
\]
Although \(D_{{\rm B},e}^{\rm LH}\) is larger than \(D_{\rm req}^{\rm LH,3.55}\), this comparison alone does not include radiative cooling. For \(B\simeq18~\mu{\rm G}\) and \(E_{e}\simeq100~{\rm TeV}\), the synchrotron cooling time is only
\[
t_{\rm syn}\sim4\times10^{2}~{\rm yr}.
\]
The associated Bohm diffusion length is of order \(\sim1~{\rm pc}\), smaller than \(d_{\rm IC}=3.55~{\rm pc}\). Thus, even in the compact lepto-hadronic case, the highest-energy electrons cannot easily propagate several parsecs in the Bohm limit without substantial cooling. Lower-energy electrons responsible for the GeV Bremsstrahlung component are less affected by these losses and can remain confined near the dense S106 gas.

We then evaluate the moderate-offset lepto-hadronic case,
\[
d_{\rm IC}=15.1~{\rm pc}\simeq15~{\rm pc}.
\]
The required diffusion coefficient becomes
\[
D_{\rm req}^{\rm LH,15}
\simeq
1.7\times10^{26}
\left(\frac{d_{\rm IC}}{15.1~{\rm pc}}\right)^2
\left(\frac{t}{10^{5}~{\rm yr}}\right)^{-1}
{\rm cm^{2}~s^{-1}} .
\]
This value is comparable to the Bohm coefficient of the \(\sim100~{\rm TeV}\) electrons. However, because the synchrotron cooling time of these electrons is only \(\sim4\times10^{2}~{\rm yr}\), their cooling-limited diffusion length remains much smaller than \(15.1~{\rm pc}\). Therefore, the offset TeV emission in the lepto-hadronic model cannot be explained by pure Bohm diffusion of the highest-energy electrons from S106 alone.

The proton component is less constrained by radiative cooling. In both lepto-hadronic transport estimates, the Bohm coefficient of the \(\sim1~{\rm PeV}\) protons is larger than the required diffusion coefficient for \(d_{\rm IC}=3.55~{\rm pc}\) and is also sufficient for transport over \(\sim15~{\rm pc}\) on timescales shorter than \(10^{5}~{\rm yr}\). Therefore, the updated lepto-hadronic solution favors a two-component interpretation: lower-energy electrons confined near the dense S106 environment produce the Fermi-LAT Bremsstrahlung emission, while the offset TeV emission requires either faster-than-Bohm diffusion in the outer region, a lower magnetic field outside the dense core, local re-acceleration, and/or a substantial hadronic contribution.

The hadronic interpretation is particularly important because the best-fit
proton cutoff reaches
\[
E_{{\rm cut},p}=1000~{\rm TeV},
\]
the upper boundary of the adopted proton energy range. Thus, the current lepto-hadronic fit supports a PeVatron-candidate interpretation. However, because the cutoff is boundary-limited, it should be treated as evidence for acceleration into the PeV regime rather than as a precisely measured spectral turnover.

%%% Conclusion
\subsection{Conclusions}
\label{sec:Concl}

In this work we have detected the LHAASO KM2A source 1LHAASO J2027+3657 with HAWC, characterizing its TeV gamma-ray emission in greater detail. We also re-analyzed \textit{Fermi}-LAT data in this part of the sky, examining especially the cataloged source 4FGL J2026.5+3718c. Combining the \textit{Fermi} and HAWC data allows for a more comprehensive understanding of the high-energy processes in this region.

We additionally re-analyzed archival \textit{Chandra} data at the star-forming region S106. Instead of focusing on the point sources within S106, which are likely young stars \cite{2004A&A...424..965G}, we extracted a spectrum for the diffuse emission in the region by masking out the point sources. We find that diffuse emission is well-fit with a $\Gamma = 2.0 \pm 0.4$ power law accounting for $\approx 10\%$ of the total X-ray luminosity of S106. This X-ray flux can be attributed to a non-thermal population of energetic particles produced in S106 by shocked stellar winds, though it may also contain unresolved point sources.

While it is often assumed that TeV emission from SFRs is hadronic in origin, our leptonic model demonstrates that the X-ray and gamma-ray emission from a star-forming region can be fully explained by leptons alone, and that gamma-rays produced by hadronic processes, do not necessarily dominate such systems. Key differences between fully leptonic and hadronic models include the morphology of the gamma-ray emission and the budget of energy in the outflows. Crucially, proper modeling of IC emission in systems like S106 must account for substantially boosted ambient photon fields that can be upscattered to TeV energies; neglecting this factor dramatically underestimates the flux of gamma-rays produced by IC scattering.

The diffusion estimates suggest a more nuanced transport picture. In the purely leptonic fit, the best-fit IC distance, \(d_{\rm IC}=3.55~{\rm pc}\), is sufficiently compact that electron transport from the S106 region remains plausible, even though the highest-energy electrons cool rapidly. In this case, the Fermi-LAT Bremsstrahlung component can be interpreted as tracing the lower-energy electron population confined in the dense S106 environment, while the IC component is produced relatively close to the acceleration site. In the lepto-hadronic interpretation, however, the higher magnetic field (\(B\simeq18~\mu{\rm G}\)) implies a short synchrotron cooling time for \(\sim100~{\rm TeV}\) electrons. Therefore, pure Bohm diffusion of the highest-energy electrons from S106 is insufficient to explain the offset TeV emission on scales of \(d_{\rm IC}\simeq15~{\rm pc}\). The TeV component may instead require faster-than-Bohm diffusion in the outer region, a lower magnetic field outside the dense core, local re-acceleration, and/or a substantial hadronic contribution. The latter possibility is particularly relevant because the lepto-hadronic fit drives the proton cutoff to \(E_{{\rm cut},p}=1000~{\rm TeV}\), supporting a PeVatron-candidate interpretation, while still leaving a degeneracy between the primary leptonic and hadronic contributions in the broadband gamma-ray SED.

The spatial coincidence of the GeV emission with the S106 SFR supports a dense-gas Bremsstrahlung origin, in which relativistic electrons accelerated by the S106IR's jet/wind system interact with the surrounding molecular material. The \textit{Fermi}-LAT emission traces the lower-energy electron population interacting with the dense gas in the S106 region. We assume these electrons to be primary particles accelerated by the S106IR's jet/wind system. However, the GeV band is intrinsically degenerate: Bremsstrahlung from primary electrons, emission from secondary \(e^\pm\) produced in hadronic interactions, and the low-energy tail of pion decay emission can contribute with partially overlapping spectral shapes. The TeV emission detected by HAWC and LHAASO probes the highest-energy particles and is modeled as a combination of IC emission from high-energy electrons and pion decay emission from accelerated protons. The fitted hadronic cutoff suggests acceleration into the multi-100~TeV regime, supporting a PeVatron-candidate interpretation, although the present data do not uniquely separate the leptonic and hadronic components.

Revisiting S106 with further X-ray observations would be a fruitful path to monitor the X-ray resolved and unresolved point sources (which are expected to have some degree of variability) and better characterize the diffuse emission (which would not be expected to vary on short timescales). It is worth noting that the gamma-ray data analyzed from \textit{Fermi}-LAT and HAWC above is comprehensive, stretching the entire durations of those observatories, while \textit{Chandra} provides only a very brief snapshot from twenty years ago.

In the future, next-generation gamma-ray observatories like the Cherenkov Telescope Array Observatory (CTAO) will have substantially greater angular resolution compared to current telescopes. This will enable spatially-resolved studies of gamma-ray emission near SFRs like S106 \citep{2026JHEAp..5000465B} for direct discrimination between the IC scattering and pion decay scenarios for gamma-ray production. The detection of neutrinos from SFRs would also be a strong discriminator between hadronic and leptonic gamma-ray production, as neutrinos are only produced during hadronic particle interactions.

\begin{acknowledgments}
This research has made use of data and/or software provided by the High Energy Astrophysics Science Archive Research Center (HEASARC), which is a service of the Astrophysics Science Division at NASA/GSFC. This work was supported by NASA through the Fermi Guest Investigator (GI) Program under grant number 80NSSC26K0546.

The HAWC Collaboration acknowledges the support from: the US National Science Foundation (NSF); the US Department of Energy Office of High-Energy Physics; the Laboratory Directed Research and Development (LDRD) program of Los Alamos National Laboratory; Consejo Nacional de Ciencia y Tecnolog\'{i}a (CONACyT), M\'{e}xico, grants LNC-2023-117, 271051, 232656, 260378, 179588, 254964, 258865, 243290, 132197, A1-S-46288, A1-S-22784, CF-2023-I-645, CBF2023-2024-1630, c\'{a}tedras 873, 1563, 341, 323, Red HAWC, M\'{e}xico; DGAPA-UNAM grants IG101323, IG100726, IN111716-3, IN111419, IA102019, IN106521, IN114924, IN110521 , IN102223; VIEP-BUAP; PIFI 2012, 2013, PROFOCIE 2014, 2015; the University of Wisconsin Alumni Research Foundation; the Institute of Geophysics, Planetary Physics, and Signatures at Los Alamos National Laboratory; Polish Science Centre grant, 2024/53/B/ST9/02671; Coordinaci\'{o}n de la Investigaci\'{o}n Cient\'{i}fica de la Universidad Michoacana; Royal Society - Newton Advanced Fellowship 180385; Gobierno de Espa\~na and European Union-NextGenerationEU, grant CNS2023- 144099; The Program Management Unit for Human Resources \& Institutional Development, Research and Innovation, NXPO (grant number B16F630069); Coordinaci\'{o}n General Acad\'{e}mica e Innovaci\'{o}n (CGAI-UdeG), PRODEP-SEP UDG-CA-499; Institute of Cosmic Ray Research (ICRR), University of Tokyo. H.M. acknowledges support under grant number CBF2023-2024-1630. H.F. acknowledges support by NASA under award number 80GSFC21M0002. C.R. acknowledges support from National Research Foundation of Korea (RS-2023-00280210, RS-2026-25470489). We also acknowledge the significant contributions over many years of Stefan Westerhoff, Gaurang Yodh and Arnulfo Zepeda Dom\'inguez, all deceased members of the HAWC collaboration. Thanks to Scott Delay, Luciano D\'{i}az and Eduardo Murrieta for technical support.
\end{acknowledgments}

%%%% Software Used
\software{3ML \citep{threeML}, \texttt{hawc-hal} \citep{hawchal}, FTools \citep{Ftools}, Xspec \citep{Xspec}, FermiPy \citep{FermiPy}, Naima \citep{naima}.}

\bibliography{main}{}

@ARTICLE{2004A&A...424..965G,
       author = {{Giardino}, G. and {Favata}, F. and {Micela}, G.},
        title = "{Chandra observations of the massive star-forming region S106. X-ray emission from the embedded massive protostellar object IRS 4}",
      journal = {\aap},
         year = 2004,
        month = sep,
       volume = {424},
        pages = {965-978},
          doi = {10.1051/0004-6361:20040363},
archivePrefix = {arXiv},
       eprint = {astro-ph/0406344},
 primaryClass = {astro-ph},
       adsurl = {https://ui.adsabs.harvard.edu/abs/2004A&A...424..965G}
}

@INPROCEEDINGS{Xspec,
       author = {{Arnaud}, K.~A.},
        title = "{XSPEC: The First Ten Years}",
    booktitle = {Astronomical Data Analysis Software and Systems V},
         year = 1996,
       editor = {{Jacoby}, George H. and {Barnes}, Jeannette},
       series = {Astronomical Society of the Pacific Conference Series},
       volume = {101},
        month = jan,
        pages = {17},
       adsurl = {https://ui.adsabs.harvard.edu/abs/1996ASPC..101...17A}
}

@software{Ftools,
       author = {{Blackburn}, J.~K. and {Shaw}, R.~A. and {Payne}, H.~E. and {Hayes}, J.~J.~E. and {Heasarc}},
        title = "{FTOOLS: A general package of software to manipulate FITS files}",
 howpublished = {Astrophysics Source Code Library, record ascl:9912.002},
         year = 1999,
        month = dec,
          eid = {ascl:9912.002},
archivePrefix = {ascl},
       eprint = {9912.002},
       adsurl = {https://ui.adsabs.harvard.edu/abs/1999ascl.soft12002B}
}

@ARTICLE{naima,
   author = {{Zabalza}, V.},
    title = {naima: a Python package for inference of relativistic particle
             energy distributions from observed nonthermal spectra},
     year = 2015,
  journal = {Proc.~of International Cosmic Ray Conference 2015},
    pages = "922",
   eprint = {1509.03319},
   adsurl = {http://adsabs.harvard.edu/abs/2015arXiv150903319Z},
}

@ARTICLE{2022Sci...378..538I,
       author = {{IceCube Collaboration} and {Abbasi}, R. and {Ackermann}, M. and {Adams}, J. and {Aguilar}, J.~A. and {Ahlers}, M. and {Ahrens}, M. and {Alameddine}, J.~M. and {Alispach}, C. and {Alves}, Jr., A.~A. and {Amin}, N.~M. and {Andeen}, K. and {Anderson}, T. and {Anton}, G. and {Arg{\"u}elles}, C. and {Ashida}, Y. and {Axani}, S. and {Bai}, X. and {Balagopal}, A.~V. and {Barbano}, V.~A. and {Barwick}, S.~W. and {Bastian}, B. and {Basu}, V. and {Baur}, S. and {Bay}, R. and {Beatty}, J.~J. and {Becker}, K. -H. and {Becker Tjus}, J. and {Bellenghi}, C. and {Benzvi}, S. and {Berley}, D. and {Bernardini}, E. and {Besson}, D.~Z. and {Binder}, G. and {Bindig}, D. and {Blaufuss}, E. and {Blot}, S. and {Boddenberg}, M. and {Bontempo}, F. and {Borowka}, J. and {B{\"o}ser}, S. and {Botner}, O. and {B{\"o}ttcher}, J. and {Bourbeau}, E. and {Bradascio}, F. and {Braun}, J. and {Brinson}, B. and {Bron}, S. and {Brostean-Kaiser}, J. and {Browne}, S. and {Burgman}, A. and {Burley}, R.~T. and {Busse}, R.~S. and {Campana}, M.~A. and {Carnie-Bronca}, E.~G. and {Chen}, C. and {Chen}, Z. and {Chirkin}, D. and {Choi}, K. and {Clark}, B.~A. and {Clark}, K. and {Classen}, L. and {Coleman}, A. and {Collin}, G.~H. and {Conrad}, J.~M. and {Coppin}, P. and {Correa}, P. and {Cowen}, D.~F. and {Cross}, R. and {Dappen}, C. and {Dave}, P. and {de Clercq}, C. and {Delaunay}, J.~J. and {Delgado L{\'o}pez}, D. and {Dembinski}, H. and {Deoskar}, K. and {Desai}, A. and {Desiati}, P. and {de Vries}, K.~D. and {de Wasseige}, G. and {de With}, M. and {Deyoung}, T. and {Diaz}, A. and {D{\'\i}az-V{\'e}lez}, J.~C. and {Dittmer}, M. and {Dujmovic}, H. and {Dunkman}, M. and {Duvernois}, M.~A. and {Dvorak}, E. and {Ehrhardt}, T. and {Eller}, P. and {Engel}, R. and {Erpenbeck}, H. and {Evans}, J. and {Evenson}, P.~A. and {Fan}, K.~L. and {Fazely}, A.~R. and {Fedynitch}, A. and {Feigl}, N. and {Fiedlschuster}, S. and {Fienberg}, A.~T. and {Filimonov}, K. and {Finley}, C. and {Fischer}, L. and {Fox}, D. and {Franckowiak}, A. and {Friedman}, E. and {Fritz}, A. and {F{\"u}rst}, P. and {Gaisser}, T.~K. and {Gallagher}, J. and {Ganster}, E. and {Garcia}, A. and {Garrappa}, S. and {Gerhardt}, L. and {Ghadimi}, A. and {Glaser}, C. and {Glauch}, T. and {Gl{\"u}senkamp}, T. and {Goldschmidt}, A. and {Gonzalez}, J.~G. and {Goswami}, S. and {Grant}, D. and {Gr{\'e}goire}, T. and {Griswold}, S. and {G{\"u}nther}, C. and {Gutjahr}, P. and {Haack}, C. and {Hallgren}, A. and {Halliday}, R. and {Halve}, L. and {Halzen}, F. and {Hanson}, M. Ha Minh K. and {Hardin}, J. and {Harnisch}, A.~A. and {Haungs}, A. and {Hebecker}, D. and {Helbing}, K. and {Henningsen}, F. and {Hettinger}, E.~C. and {Hickford}, S. and {Hignight}, J. and {Hill}, C. and {Hill}, G.~C. and {Hoffman}, K.~D. and {Hoffmann}, R. and {Hokanson-Fasig}, B. and {Hoshina}, K. and {Huang}, F. and {Huber}, M. and {Huber}, T. and {Hultqvist}, K. and {H{\"u}nnefeld}, M. and {Hussain}, R. and {Hymon}, K. and {in}, S. and {Iovine}, N. and {Ishihara}, A. and {Jansson}, M. and {Japaridze}, G.~S. and {Jeong}, M. and {Jin}, M. and {Jones}, B.~J.~P. and {Kang}, D. and {Kang}, W. and {Kang}, X. and {Kappes}, A. and {Kappesser}, D. and {Kardum}, L. and {Karg}, T. and {Karl}, M. and {Karle}, A. and {Katz}, U. and {Kauer}, M. and {Kellermann}, M. and {Kelley}, J.~L. and {Kheirandish}, A. and {Kin}, K. and {Kintscher}, T. and {Kiryluk}, J. and {Klein}, S.~R. and {Koirala}, R. and {Kolanoski}, H. and {Kontrimas}, T. and {K{\"o}pke}, L. and {Kopper}, C. and {Kopper}, S. and {Koskinen}, D.~J. and {Koundal}, P. and {Kovacevich}, M. and {Kowalski}, M. and {Kozynets}, T. and {Kun}, E. and {Kurahashi}, N. and {Lad}, N. and {Lagunas Gualda}, C. and {Lanfranchi}, J.~L. and {Larson}, M.~J. and {Lauber}, F. and {Lazar}, J.~P.},
        title = "{Evidence for neutrino emission from the nearby active galaxy NGC 1068}",
      journal = {Science},
         year = 2022,
        month = nov,
       volume = {378},
       number = {6619},
        pages = {538-543},
          doi = {10.1126/science.abg3395},
archivePrefix = {arXiv},
       eprint = {2211.09972},
 primaryClass = {astro-ph.HE},
       adsurl = {https://ui.adsabs.harvard.edu/abs/2022Sci...378..538I}
}

@article{2023Sci...380.1338I,
author = {IceCube Collaboration*† and R. Abbasi  and M. Ackermann  and J. Adams  and J. A. Aguilar  and M. Ahlers  and M. Ahrens  and J. M. Alameddine  and A. A. Alves  and N. M. Amin  and K. Andeen  and T. Anderson  and G. Anton  and C. Argüelles  and Y. Ashida  and S. Athanasiadou  and S. Axani  and X. Bai  and A. Balagopal V.  and S. W. Barwick  and V. Basu  and S. Baur  and R. Bay  and J. J. Beatty  and K.-H. Becker  and J. Becker Tjus  and J. Beise  and C. Bellenghi  and S. Benda  and S. BenZvi  and D. Berley  and E. Bernardini  and D. Z. Besson  and G. Binder  and D. Bindig  and E. Blaufuss  and S. Blot  and M. Boddenberg  and F. Bontempo  and J. Y. Book  and J. Borowka  and S. Böser  and O. Botner  and J. Böttcher  and E. Bourbeau  and F. Bradascio  and J. Braun  and B. Brinson  and S. Bron  and J. Brostean-Kaiser  and R. T. Burley  and R. S. Busse  and M. A. Campana  and E. G. Carnie-Bronca  and C. Chen  and Z. Chen  and D. Chirkin  and K. Choi  and B. A. Clark  and K. Clark  and L. Classen  and A. Coleman  and G. H. Collin  and A. Connolly  and J. M. Conrad  and P. Coppin  and P. Correa  and D. F. Cowen  and R. Cross  and C. Dappen  and P. Dave  and C. De Clercq  and J. J. DeLaunay  and D. Delgado López  and H. Dembinski  and K. Deoskar  and A. Desai  and P. Desiati  and K. D. de Vries  and G. de Wasseige  and T. DeYoung  and A. Diaz  and J. C. Díaz-Vélez  and M. Dittmer  and H. Dujmovic  and M. Dunkman  and M. A. DuVernois  and T. Ehrhardt  and P. Eller  and R. Engel  and H. Erpenbeck  and J. Evans  and P. A. Evenson  and K. L. Fan  and A. R. Fazely  and A. Fedynitch  and N. Feigl  and S. Fiedlschuster  and A. T. Fienberg  and C. Finley  and L. Fischer  and D. Fox  and A. Franckowiak  and E. Friedman  and A. Fritz  and P. Fürst  and T. K. Gaisser  and J. Gallagher  and E. Ganster  and A. Garcia  and S. Garrappa  and L. Gerhardt  and A. Ghadimi  and C. Glaser  and T. Glauch  and T. Glüsenkamp  and N. Goehlke  and A. Goldschmidt  and J. G. Gonzalez  and S. Goswami  and D. Grant  and T. Grégoire  and S. Griswold  and C. Günther  and P. Gutjahr  and C. Haack  and A. Hallgren  and R. Halliday  and L. Halve  and F. Halzen  and M. Ha Minh  and K. Hanson  and J. Hardin  and A. A. Harnisch  and A. Haungs  and K. Helbing  and F. Henningsen  and E. C. Hettinger  and S. Hickford  and J. Hignight  and C. Hill  and G. C. Hill  and K. D. Hoffman  and K. Hoshina  and W. Hou  and F. Huang  and M. Huber  and T. Huber  and K. Hultqvist  and M. Hünnefeld  and R. Hussain  and K. Hymon  and S. In  and N. Iovine  and A. Ishihara  and M. Jansson  and G. S. Japaridze  and M. Jeong  and M. Jin  and B. J. P. Jones  and D. Kang  and W. Kang  and X. Kang  and A. Kappes  and D. Kappesser  and L. Kardum  and T. Karg  and M. Karl  and A. Karle  and U. Katz  and M. Kauer  and M. Kellermann  and J. L. Kelley  and A. Kheirandish  and K. Kin  and J. Kiryluk  and S. R. Klein  and A. Kochocki  and R. Koirala  and H. Kolanoski  and T. Kontrimas  and L. Köpke  and C. Kopper  and S. Kopper  and D. J. Koskinen  and P. Koundal  and M. Kovacevich  and M. Kowalski  and T. Kozynets  and E. Krupczak  and E. Kun  and N. Kurahashi  and N. Lad  and C. Lagunas Gualda  and J. L. Lanfranchi  and M. J. Larson  and F. Lauber  and J. P. Lazar  and J. W. Lee  and K. Leonard  and A. Leszczyńska  and Y. Li  and M. Lincetto  and Q. R. Liu  and M. Liubarska  and E. Lohfink  and C. J. Lozano Mariscal  and L. Lu  and F. Lucarelli  and A. Ludwig  and W. Luszczak  and Y. Lyu  and W. Y. Ma  and J. Madsen  and K. B. M. Mahn  and Y. Makino  and S. Mancina  and I. C. Mariş  and I. Martinez-Soler  and R. Maruyama  and S. McHale  and T. McElroy  and F. McNally  and J. V. Mead  and K. Meagher  and S. Mechbal  and A. Medina  and M. Meier  and S. Meighen-Berger  and Y. Merckx  and J. Micallef  and D. Mockler  and T. Montaruli  and R. W. Moore  and K. Morik  and R. Morse  and M. Moulai  and T. Mukherjee  and R. Naab  and R. Nagai  and R. Nahnhauer  and U. Naumann  and J. Necker  and L. V. Nguyen  and H. Niederhausen  and M. U. Nisa  and S. C. Nowicki  and D. Nygren  and A. Obertacke Pollmann  and M. Oehler  and B. Oeyen  and A. Olivas  and E. O'Sullivan  and H. Pandya  and D. V. Pankova  and N. Park  and G. K. Parker  and E. N. Paudel  and L. Paul  and C. Pérez de los Heros  and L. Peters  and J. Peterson  and S. Philippen  and S. Pieper  and A. Pizzuto  and M. Plum  and Y. Popovych  and A. Porcelli  and M. Prado Rodriguez  and B. Pries  and G. T. Przybylski  and C. Raab  and J. Rack-Helleis  and A. Raissi  and M. Rameez  and K. Rawlins  and I. C. Rea  and Z. Rechav  and A. Rehman  and P. Reichherzer  and R. Reimann  and G. Renzi  and E. Resconi  and S. Reusch  and W. Rhode  and M. Richman  and B. Riedel  and E. J. Roberts  and S. Robertson  and G. Roellinghoff  and M. Rongen  and C. Rott  and T. Ruhe  and D. Ryckbosch  and D. Rysewyk Cantu  and I. Safa  and J. Saffer  and D. Salazar-Gallegos  and P. Sampathkumar  and S. E. Sanchez Herrera  and A. Sandrock  and M. Santander  and S. Sarkar  and S. Sarkar  and K. Satalecka  and M. Schaufel  and H. Schieler  and S. Schindler  and T. Schmidt  and A. Schneider  and J. Schneider  and F. G. Schröder  and L. Schumacher  and G. Schwefer  and S. Sclafani  and D. Seckel  and S. Seunarine  and A. Sharma  and S. Shefali  and N. Shimizu  and M. Silva  and B. Skrzypek  and B. Smithers  and R. Snihur  and J. Soedingrekso  and A. Sogaard  and D. Soldin  and C. Spannfellner  and G. M. Spiczak  and C. Spiering  and M. Stamatikos  and T. Stanev  and R. Stein  and J. Stettner  and T. Stezelberger  and B. Stokstad  and T. Stürwald  and T. Stuttard  and G. W. Sullivan  and I. Taboada  and S. Ter-Antonyan  and J. Thwaites  and S. Tilav  and F. Tischbein  and K. Tollefson  and C. Tönnis  and S. Toscano  and D. Tosi  and A. Trettin  and M. Tselengidou  and C. F. Tung  and A. Turcati  and R. Turcotte  and C. F. Turley  and J. P. Twagirayezu  and B. Ty  and M. A. Unland Elorrieta  and N. Valtonen-Mattila  and J. Vandenbroucke  and N. van Eijndhoven  and D. Vannerom  and J. van Santen  and J. Veitch-Michaelis  and S. Verpoest  and C. Walck  and W. Wang  and T. B. Watson  and C. Weaver  and P. Weigel  and A. Weindl  and M. J. Weiss  and J. Weldert  and C. Wendt  and J. Werthebach  and M. Weyrauch  and N. Whitehorn  and C. H. Wiebusch  and N. Willey  and D. R. Williams  and M. Wolf  and G. Wrede  and J. Wulff  and X. W. Xu  and J. P. Yanez  and E. Yildizci  and S. Yoshida  and S. Yu  and T. Yuan  and Z. Zhang  and P. Zhelnin },
title = {Observation of high-energy neutrinos from the Galactic plane},
journal = {Science},
volume = {380},
number = {6652},
pages = {1338-1343},
year = {2023},
doi = {10.1126/science.adc9818},
URL = {https://www.science.org/doi/abs/10.1126/science.adc9818},
eprint = {https://www.science.org/doi/pdf/10.1126/science.adc9818}}

@ARTICLE{2018Sci...361..147I,
       author = {{IceCube Collaboration} and {Aartsen}, M.~G. and {Ackermann}, M. and {Adams}, J. and {Aguilar}, J.~A. and {Ahlers}, M. and {Ahrens}, M. and {Samarai}, I. Al and {Altmann}, D. and {Andeen}, K. and {Anderson}, T. and {Ansseau}, I. and {Anton}, G. and {Arg{\"u}elles}, C. and {Arsioli}, B. and {Auffenberg}, J. and {Axani}, S. and {Bagherpour}, H. and {Bai}, X. and {Barron}, J.~P. and {Barwick}, S.~W. and {Baum}, V. and {Bay}, R. and {Beatty}, J.~J. and {Becker Tjus}, J. and {Becker}, K. -H. and {BenZvi}, S. and {Berley}, D. and {Bernardini}, E. and {Besson}, D.~Z. and {Binder}, G. and {Bindig}, D. and {Blaufuss}, E. and {Blot}, S. and {Bohm}, C. and {B{\"o}rner}, M. and {Bos}, F. and {B{\"o}ser}, S. and {Botner}, O. and {Bourbeau}, E. and {Bourbeau}, J. and {Bradascio}, F. and {Braun}, J. and {Brenzke}, M. and {Bretz}, H. -P. and {Bron}, S. and {Brostean-Kaiser}, J. and {Burgman}, A. and {Busse}, R.~S. and {Carver}, T. and {Cheung}, E. and {Chirkin}, D. and {Christov}, A. and {Clark}, K. and {Classen}, L. and {Coenders}, S. and {Collin}, G.~H. and {Conrad}, J.~M. and {Coppin}, P. and {Correa}, P. and {Cowen}, D.~F. and {Cross}, R. and {Dave}, P. and {Day}, M. and {de Andr{\'e}}, J.~P.~A.~M. and {De Clercq}, C. and {DeLaunay}, J.~J. and {Dembinski}, H. and {DeRidder}, S. and {Desiati}, P. and {de Vries}, K.~D. and {de Wasseige}, G. and {de With}, M. and {DeYoung}, T. and {D{\'\i}az-V{\'e}lez}, J.~C. and {di Lorenzo}, V. and {Dujmovic}, H. and {Dumm}, J.~P. and {Dunkman}, M. and {Dvorak}, E. and {Eberhardt}, B. and {Ehrhardt}, T. and {Eichmann}, B. and {Eller}, P. and {Evenson}, P.~A. and {Fahey}, S. and {Fazely}, A.~R. and {Felde}, J. and {Filimonov}, K. and {Finley}, C. and {Flis}, S. and {Franckowiak}, A. and {Friedman}, E. and {Fritz}, A. and {Gaisser}, T.~K. and {Gallagher}, J. and {Gerhardt}, L. and {Ghorbani}, K. and {Giommi}, P. and {Glauch}, T. and {Gl{\"u}senkamp}, T. and {Goldschmidt}, A. and {Gonzalez}, J.~G. and {Grant}, D. and {Griffith}, Z. and {Haack}, C. and {Hallgren}, A. and {Halzen}, F. and {Hanson}, K. and {Hebecker}, D. and {Heereman}, D. and {Helbing}, K. and {Hellauer}, R. and {Hickford}, S. and {Hignight}, J. and {Hill}, G.~C. and {Hoffman}, K.~D. and {Hoffmann}, R. and {Hoinka}, T. and {Hokanson-Fasig}, B. and {Hoshina}, K. and {Huang}, F. and {Huber}, M. and {Hultqvist}, K. and {H{\"u}nnefeld}, M. and {Hussain}, R. and {In}, S. and {Iovine}, N. and {Ishihara}, A. and {Jacobi}, E. and {Japaridze}, G.~S. and {Jeong}, M. and {Jero}, K. and {Jones}, B.~J.~P. and {Kalaczynski}, P. and {Kang}, W. and {Kappes}, A. and {Kappesser}, D. and {Karg}, T. and {Karle}, A. and {Katz}, U. and {Kauer}, M. and {Keivani}, A. and {Kelley}, J.~L. and {Kheirandish}, A. and {Kim}, J. and {Kim}, M. and {Kintscher}, T. and {Kiryluk}, J. and {Kittler}, T. and {Klein}, S.~R. and {Koirala}, R. and {Kolanoski}, H. and {K{\"o}pke}, L. and {Kopper}, C. and {Kopper}, S. and {Koschinsky}, J.~P. and {Koskinen}, D.~J. and {Kowalski}, M. and {Krammer}, B. and {Krings}, K. and {Kroll}, M. and {Kr{\"u}ckl}, G. and {Kunwar}, S. and {Kurahashi}, N. and {Kuwabara}, T. and {Kyriacou}, A. and {Labare}, M. and {Lanfranchi}, J.~L. and {Larson}, M.~J. and {Lauber}, F. and {Leonard}, K. and {Lesiak-Bzdak}, M. and {Leuermann}, M. and {Liu}, Q.~R. and {Lozano Mariscal}, C.~J. and {Lu}, L. and {L{\"u}nemann}, J. and {Luszczak}, W. and {Madsen}, J. and {Maggi}, G. and {Mahn}, K.~B.~M. and {Mancina}, S. and {Maruyama}, R. and {Mase}, K. and {Maunu}, R. and {Meagher}, K. and {Medici}, M. and {Meier}, M. and {Menne}, T. and {Merino}, G. and {Meures}, T. and {Miarecki}, S. and {Micallef}, J. and {Moment{\'e}}, G. and {Montaruli}, T. and {Moore}, R.~W. and {Morse}, R. and {Moulai}, M. and {Nahnhauer}, R.},
        title = "{Neutrino emission from the direction of the blazar TXS 0506+056 prior to the IceCube-170922A alert}",
      journal = {Science},
         year = 2018,
        month = jul,
       volume = {361},
       number = {6398},
        pages = {147-151},
          doi = {10.1126/science.aat2890},
archivePrefix = {arXiv},
       eprint = {1807.08794},
 primaryClass = {astro-ph.HE},
       adsurl = {https://ui.adsabs.harvard.edu/abs/2018Sci...361..147I}
}

@article{2025arXiv250604109L,
    author = "Lien, Amy Y. and others",
    title = "{The 157 Month Swift/BAT All-sky Hard X-Ray Survey}",
    eprint = "2506.04109",
    archivePrefix = "arXiv",
    primaryClass = "astro-ph.HE",
    doi = "10.3847/1538-4357/ade676",
    journal = "Astrophys. J.",
    volume = "989",
    number = "2",
    pages = "161",
    year = "2025"
}

@ARTICLE{2005ApJ...627..263V,
       author = {{Vall{\'e}e}, Jacques P. and {Fiege}, Jason D.},
        title = "{A Cool Magnetized Shell Wrapped around the Hot H II Region S106: Geometry, Kinematics, Magnetic Vectors, and Pressure Balance}",
      journal = {\apj},
         year = 2005,
        month = jul,
       volume = {627},
       number = {1},
        pages = {263-276},
          doi = {10.1086/430341},
       adsurl = {https://ui.adsabs.harvard.edu/abs/2005ApJ...627..263V}
}

@ARTICLE{1995ApJ...442..208R,
       author = {{Roberts}, D.~A. and {Crutcher}, R.~M. and {Troland}, T.~H.},
        title = "{The Distribution of Molecular and Neutral Gas and Magnetic Fields near the Bipolar H II Region S106}",
      journal = {\apj},
         year = 1995,
        month = mar,
       volume = {442},
        pages = {208},
          doi = {10.1086/175436},
       adsurl = {https://ui.adsabs.harvard.edu/abs/1995ApJ...442..208R}
}

@ARTICLE{1988ApJ...335..263K,
       author = {{Kazes}, I. and {Troland}, T.~H. and {Crutcher}, R.~M. and {Heiles}, C.},
        title = "{The Magnetic Field in the Vicinity of S106}",
      journal = {\apj},
         year = 1988,
        month = dec,
       volume = {335},
        pages = {263},
          doi = {10.1086/166924},
       adsurl = {https://ui.adsabs.harvard.edu/abs/1988ApJ...335..263K}
}

@ARTICLE{2024ApJS..271...25C,
       author = {{Cao}, Zhen and {Aharonian}, F. and {An}, Q. and {Axikegu} and {Bai}, Y.~X. and {Bao}, Y.~W. and {Bastieri}, D. and {Bi}, X.~J. and {Bi}, Y.~J. and {Cai}, J.~T. and {Cao}, Q. and {Cao}, W.~Y. and {Cao}, Zhe and {Chang}, J. and {Chang}, J.~F. and {Chen}, A.~M. and {Chen}, E.~S. and {Chen}, Liang and {Chen}, Lin and {Chen}, Long and {Chen}, M.~J. and {Chen}, M.~L. and {Chen}, Q.~H. and {Chen}, S.~H. and {Chen}, S.~Z. and {Chen}, T.~L. and {Chen}, Y. and {Cheng}, N. and {Cheng}, Y.~D. and {Cui}, M.~Y. and {Cui}, S.~W. and {Cui}, X.~H. and {Cui}, Y.~D. and {Dai}, B.~Z. and {Dai}, H.~L. and {Dai}, Z.~G. and {Danzengluobu} and {Della Volpe}, D. and {Dong}, X.~Q. and {Duan}, K.~K. and {Fan}, J.~H. and {Fan}, Y.~Z. and {Fang}, J. and {Fang}, K. and {Feng}, C.~F. and {Feng}, L. and {Feng}, S.~H. and {Feng}, X.~T. and {Feng}, Y.~L. and {Gabici}, S. and {Gao}, B. and {Gao}, C.~D. and {Gao}, L.~Q. and {Gao}, Q. and {Gao}, W. and {Gao}, W.~K. and {Ge}, M.~M. and {Geng}, L.~S. and {Giacinti}, G. and {Gong}, G.~H. and {Gou}, Q.~B. and {Gu}, M.~H. and {Guo}, F.~L. and {Guo}, X.~L. and {Guo}, Y.~Q. and {Guo}, Y.~Y. and {Han}, Y.~A. and {He}, H.~H. and {He}, H.~N. and {He}, J.~Y. and {He}, X.~B. and {He}, Y. and {Heller}, M. and {Hor}, Y.~K. and {Hou}, B.~W. and {Hou}, C. and {Hou}, X. and {Hu}, H.~B. and {Hu}, Q. and {Hu}, S.~C. and {Huang}, D.~H. and {Huang}, T.~Q. and {Huang}, W.~J. and {Huang}, X.~T. and {Huang}, X.~Y. and {Huang}, Y. and {Huang}, Z.~C. and {Ji}, X.~L. and {Jia}, H.~Y. and {Jia}, K. and {Jiang}, K. and {Jiang}, X.~W. and {Jiang}, Z.~J. and {Jin}, M. and {Kang}, M.~M. and {Ke}, T. and {Kuleshov}, D. and {Kurinov}, K. and {Li}, B.~B. and {Li}, Cheng and {Li}, Cong and {Li}, D. and {Li}, F. and {Li}, H.~B. and {Li}, H.~C. and {Li}, H.~Y. and {Li}, J. and {Li}, Jian and {Li}, Jie and {Li}, K. and {Li}, W.~L. and {Li}, W.~L. and {Li}, X.~R. and {Li}, Xin and {Li}, Y.~Z. and {Li}, Zhe and {Li}, Zhuo and {Liang}, E.~W. and {Liang}, Y.~F. and {Lin}, S.~J. and {Liu}, B. and {Liu}, C. and {Liu}, D. and {Liu}, H. and {Liu}, H.~D. and {Liu}, J. and {Liu}, J.~L. and {Liu}, J.~Y. and {Liu}, M.~Y. and {Liu}, R.~Y. and {Liu}, S.~M. and {Liu}, W. and {Liu}, Y. and {Liu}, Y.~N. and {Lu}, R. and {Luo}, Q. and {Lv}, H.~K. and {Ma}, B.~Q. and {Ma}, L.~L. and {Ma}, X.~H. and {Mao}, J.~R. and {Min}, Z. and {Mitthumsiri}, W. and {Mu}, H.~J. and {Nan}, Y.~C. and {Neronov}, A. and {Ou}, Z.~W. and {Pang}, B.~Y. and {Pattarakijwanich}, P. and {Pei}, Z.~Y. and {Qi}, M.~Y. and {Qi}, Y.~Q. and {Qiao}, B.~Q. and {Qin}, J.~J. and {Ruffolo}, D. and {S{\'a}iz}, A. and {Semikoz}, D. and {Shao}, C.~Y. and {Shao}, L. and {Shchegolev}, O. and {Sheng}, X.~D. and {Shu}, F.~W. and {Song}, H.~C. and {Stenkin}, Yu. V. and {Stepanov}, V. and {Su}, Y. and {Sun}, Q.~N. and {Sun}, X.~N. and {Sun}, Z.~B. and {Tam}, P.~H.~T. and {Tang}, Q.~W. and {Tang}, Z.~B. and {Tian}, W.~W. and {Wang}, C. and {Wang}, C.~B. and {Wang}, G.~W. and {Wang}, H.~G. and {Wang}, H.~H. and {Wang}, J.~C. and {Wang}, K. and {Wang}, L.~P. and {Wang}, L.~Y. and {Wang}, P.~H. and {Wang}, R. and {Wang}, W. and {Wang}, X.~G. and {Wang}, X.~Y. and {Wang}, Y. and {Wang}, Y.~D. and {Wang}, Y.~J. and {Wang}, Z.~H. and {Wang}, Z.~X. and {Wang}, Zhen and {Wang}, Zheng and {Wei}, D.~M. and {Wei}, J.~J. and {Wei}, Y.~J. and {Wen}, T. and {Wu}, C.~Y. and {Wu}, H.~R.},
        title = "{The First LHAASO Catalog of Gamma-Ray Sources}",
      journal = {\apjs},
         year = 2024,
        month = mar,
       volume = {271},
       number = {1},
          eid = {25},
        pages = {25},
          doi = {10.3847/1538-4365/acfd29},
archivePrefix = {arXiv},
       eprint = {2305.17030},
 primaryClass = {astro-ph.HE},
       adsurl = {https://ui.adsabs.harvard.edu/abs/2024ApJS..271...25C}
}

@article{Comeron_2018,
       author = {{Comer{\'o}n}, F. and {Schneider}, N. and {Djupvik}, A.~A. and {Schnugg}, C.},
        title = "{The ionizing source of the bipolar HII region S106: A close massive binary}",
      journal = {\aap},
         year = 2018,
        month = jul,
       volume = {615},
          eid = {A2},
        pages = {A2},
          doi = {10.1051/0004-6361/201731979},
archivePrefix = {arXiv},
       eprint = {1801.08958},
 primaryClass = {astro-ph.SR},
       adsurl = {https://ui.adsabs.harvard.edu/abs/2018A&A...615A...2C}
}

@INPROCEEDINGS{FermiPy,
       author = {{Wood}, M. and {Caputo}, R. and {Charles}, E. and {Di Mauro}, M. and {Magill}, J. and {Perkins}, J.~S. and {Fermi-LAT Collaboration}},
        title = "{Fermipy: An open-source Python package for analysis of Fermi-LAT Data}",
    booktitle = {35th International Cosmic Ray Conference (ICRC2017)},
         year = 2017,
       series = {International Cosmic Ray Conference},
       volume = {301},
        month = jul,
          eid = {824},
        pages = {824},
          doi = {10.22323/1.301.0824},
archivePrefix = {arXiv},
       eprint = {1707.09551},
 primaryClass = {astro-ph.IM},
       adsurl = {https://ui.adsabs.harvard.edu/abs/2017ICRC...35..824W}
}

@ARTICLE{4FGLDR4,
       author = {{Ballet}, J. and {Bruel}, P. and {Burnett}, T.~H. and {Lott}, B. and {The Fermi-LAT collaboration}},
        title = "{Fermi Large Area Telescope Fourth Source Catalog Data Release 4 (4FGL-DR4)}",
      journal = {arXiv e-prints},
         year = 2023,
        month = jul,
          eid = {arXiv:2307.12546},
        pages = {arXiv:2307.12546},
          doi = {10.48550/arXiv.2307.12546},
archivePrefix = {arXiv},
       eprint = {2307.12546},
 primaryClass = {astro-ph.HE},
       adsurl = {https://ui.adsabs.harvard.edu/abs/2023arXiv230712546B}
}

@misc{LATmodels,
  author = {GSFC},
  year = {2024},
  title = {Fermi-LAT Background Models},
  howpublished = {\url{https://fermi.gsfc.nasa.gov/ssc/data/access/lat/BackgroundModels.html}},
  note = {Accessed: 2024-11-11}
}

@ARTICLE{2021MNRAS.504.1840C,
       author = {{Collins}, T. and {Rowell}, G. and {Mitchell}, A.~M.~W. and {Voisin}, F. and {Fukui}, Y. and {Sano}, H. and {Alsulami}, R. and {Einecke}, S.},
        title = "{Explaining the extended GeV gamma-ray emission adjacent to HESS J1825-137}",
      journal = {\mnras},
         year = 2021,
        month = jun,
       volume = {504},
       number = {2},
        pages = {1840-1853},
          doi = {10.1093/mnras/stab983},
archivePrefix = {arXiv},
       eprint = {2104.05953},
 primaryClass = {astro-ph.HE},
       adsurl = {https://ui.adsabs.harvard.edu/abs/2021MNRAS.504.1840C}
}

@ARTICLE{2014PhRvD..90b3526A,
       author = {{Abazajian}, Kevork N. and {Canac}, Nicolas and {Horiuchi}, Shunsaku and {Kaplinghat}, Manoj},
        title = "{Astrophysical and dark matter interpretations of extended gamma-ray emission from the Galactic Center}",
      journal = {\prd},
         year = 2014,
        month = jul,
       volume = {90},
       number = {2},
          eid = {023526},
        pages = {023526},
          doi = {10.1103/PhysRevD.90.023526},
archivePrefix = {arXiv},
       eprint = {1402.4090},
 primaryClass = {astro-ph.HE},
       adsurl = {https://ui.adsabs.harvard.edu/abs/2014PhRvD..90b3526A}
}

@ARTICLE{Ro10,
       author = {{Romero}, Gustavo E.},
        title = "{Gamma-ray emission from star-forming regions.}",
      journal = {\memsai},
         year = 2010,
        month = jan,
       volume = {81},
        pages = {181},
       adsurl = {https://ui.adsabs.harvard.edu/abs/2010MmSAI..81..181R}
}

@article{Er20,
    author = {Ergin, T and Saha, L and Bhattacharjee, P and Sano, H and Tanaka, S J and Majumdar, P and Yamazaki, R and Fukui, Y},
    title = {Probing the star formation origin of gamma-rays from 3FHL J1907.0+0713},
    journal = {Monthly Notices of the Royal Astronomical Society},
    volume = {501},
    number = {3},
    pages = {4226-4237},
    year = {2020},
    month = {12},
    issn = {0035-8711},
    doi = {10.1093/mnras/staa3817},
    url = {https://doi.org/10.1093/mnras/staa3817},
    eprint = {https://academic.oup.com/mnras/article-pdf/501/3/4226/35873714/staa3817.pdf},
}

@ARTICLE{Re15,
       author = {{Reitberger}, K. and {Reimer}, A. and {Reimer}, O. and {Takahashi}, H.},
        title = "{The first full orbit of {\ensuremath{\eta}} Carinae seen by Fermi}",
      journal = {\aap},
         year = 2015,
        month = may,
       volume = {577},
          eid = {A100},
        pages = {A100},
          doi = {10.1051/0004-6361/201525726},
archivePrefix = {arXiv},
       eprint = {1503.07637},
 primaryClass = {astro-ph.HE},
       adsurl = {https://ui.adsabs.harvard.edu/abs/2015A&A...577A.100R}
}

@ARTICLE{Ta09,
       author = {{Tavani}, M. and {Sabatini}, S. and {Pian}, E. and {Bulgarelli}, A. and {Caraveo}, P. and {Viotti}, R.~F. and {Corcoran}, M.~F. and {Giuliani}, A. and {Pittori}, C. and {Verrecchia}, F. and {Vercellone}, S. and {Mereghetti}, S. and {Argan}, A. and {Barbiellini}, G. and {Boffelli}, F. and {Cattaneo}, P.~W. and {Chen}, A.~W. and {Cocco}, V. and {D'Ammando}, F. and {Costa}, E. and {DeParis}, G. and {Del Monte}, E. and {Di Cocco}, G. and {Donnarumma}, I. and {Evangelista}, Y. and {Ferrari}, A. and {Feroci}, M. and {Fiorini}, M. and {Froysland}, T. and {Fuschino}, F. and {Galli}, M. and {Gianotti}, F. and {Labanti}, C. and {Lapshov}, I. and {Lazzarotto}, F. and {Lipari}, P. and {Longo}, F. and {Marisaldi}, M. and {Mastropietro}, M. and {Morelli}, E. and {Moretti}, E. and {Morselli}, A. and {Pacciani}, L. and {Pellizzoni}, A. and {Perotti}, F. and {Piano}, G. and {Picozza}, P. and {Pilia}, M. and {Porrovecchio}, G. and {Pucella}, G. and {Prest}, M. and {Rapisarda}, M. and {Rappoldi}, A. and {Rubini}, A. and {Soffitta}, P. and {Trifoglio}, M. and {Trois}, A. and {Vallazza}, E. and {Vittorini}, V. and {Zambra}, A. and {Zanello}, D. and {Santolamazza}, P. and {Giommi}, P. and {Colafrancesco}, S. and {Antonelli}, L.~A. and {Salotti}, L.},
        title = "{Detection of Gamma-Ray Emission from the Eta-Carinae Region}",
      journal = {\apjl},
         year = 2009,
        month = jun,
       volume = {698},
       number = {2},
        pages = {L142-L146},
          doi = {10.1088/0004-637X/698/2/L142},
archivePrefix = {arXiv},
       eprint = {0904.2736},
 primaryClass = {astro-ph.HE},
       adsurl = {https://ui.adsabs.harvard.edu/abs/2009ApJ...698L.142T}
}

@ARTICLE{Be11,
       author = {{Bednarek}, W. and {Pabich}, J.},
        title = "{High-energy radiation from the massive binary system Eta Carinae}",
      journal = {\aap},
         year = 2011,
        month = jun,
       volume = {530},
          eid = {A49},
        pages = {A49},
          doi = {10.1051/0004-6361/201116549},
archivePrefix = {arXiv},
       eprint = {1104.1275},
 primaryClass = {astro-ph.HE},
       adsurl = {https://ui.adsabs.harvard.edu/abs/2011A&A...530A..49B}
}

@ARTICLE{Ps16,
       author = {{Pshirkov}, M.~S.},
        title = "{The Fermi-LAT view of the colliding wind binaries}",
      journal = {\mnras},
         year = 2016,
        month = mar,
       volume = {457},
       number = {1},
        pages = {L99-L102},
          doi = {10.1093/mnrasl/slv205},
archivePrefix = {arXiv},
       eprint = {1510.03885},
 primaryClass = {astro-ph.HE},
       adsurl = {https://ui.adsabs.harvard.edu/abs/2016MNRAS.457L..99P}
}

@ARTICLE{Ah19,
       author = {{Aharonian}, Felix and {Yang}, Ruizhi and {de O{\~n}a Wilhelmi}, Emma},
        title = "{Massive stars as major factories of Galactic cosmic rays}",
      journal = {Nature Astronomy},
         year = 2019,
        month = mar,
       volume = {3},
        pages = {561-567},
          doi = {10.1038/s41550-019-0724-0},
archivePrefix = {arXiv},
       eprint = {1804.02331},
 primaryClass = {astro-ph.HE},
       adsurl = {https://ui.adsabs.harvard.edu/abs/2019NatAs...3..561A}
}

@ARTICLE{Sa20,
       author = {{Saha}, L. and {Dom{\'\i}nguez}, A. and {Tibaldo}, L. and {Marchesi}, S. and {Ajello}, M. and {Lemoine-Goumard}, M. and {L{\'o}pez}, M.},
        title = "{Morphological and Spectral Study of 4FGL J1115.1-6118 in the Region of the Young Massive Stellar Cluster NGC 3603}",
      journal = {\apj},
         year = 2020,
        month = jul,
       volume = {897},
       number = {2},
          eid = {131},
        pages = {131},
          doi = {10.3847/1538-4357/ab9ac2},
archivePrefix = {arXiv},
       eprint = {2006.00274},
 primaryClass = {astro-ph.HE},
       adsurl = {https://ui.adsabs.harvard.edu/abs/2020ApJ...897..131S}
}

@ARTICLE{Ac11,
       author = {{Ackermann}, M. and {Ajello}, M. and {Allafort}, A. and {Baldini}, L. and {Ballet}, J. and {Barbiellini}, G. and {Bastieri}, D. and {Belfiore}, A. and {Bellazzini}, R. and {Berenji}, B. and {Blandford}, R.~D. and {Bloom}, E.~D. and {Bonamente}, E. and {Borgland}, A.~W. and {Bottacini}, E. and {Brigida}, M. and {Bruel}, P. and {Buehler}, R. and {Buson}, S. and {Caliandro}, G.~A. and {Cameron}, R.~A. and {Caraveo}, P.~A. and {Casandjian}, J.~M. and {Cecchi}, C. and {Chekhtman}, A. and {Cheung}, C.~C. and {Chiang}, J. and {Ciprini}, S. and {Claus}, R. and {Cohen-Tanugi}, J. and {de Angelis}, A. and {de Palma}, F. and {Dermer}, C.~D. and {do Couto e Silva}, E. and {Drell}, P.~S. and {Dumora}, D. and {Favuzzi}, C. and {Fegan}, S.~J. and {Focke}, W.~B. and {Fortin}, P. and {Fukazawa}, Y. and {Fusco}, P. and {Gargano}, F. and {Germani}, S. and {Giglietto}, N. and {Giordano}, F. and {Giroletti}, M. and {Glanzman}, T. and {Godfrey}, G. and {Grenier}, I.~A. and {Guillemot}, L. and {Guiriec}, S. and {Hadasch}, D. and {Hanabata}, Y. and {Harding}, A.~K. and {Hayashida}, M. and {Hayashi}, K. and {Hays}, E. and {J{\'o}hannesson}, G. and {Johnson}, A.~S. and {Kamae}, T. and {Katagiri}, H. and {Kataoka}, J. and {Kerr}, M. and {Kn{\"o}dlseder}, J. and {Kuss}, M. and {Lande}, J. and {Latronico}, L. and {Lee}, S.-H. and {Longo}, F. and {Loparco}, F. and {Lott}, B. and {Lovellette}, M.~N. and {Lubrano}, P. and {Martin}, P. and {Mazziotta}, M.~N. and {McEnery}, J.~E. and {Mehault}, J. and {Michelson}, P.~F. and {Mitthumsiri}, W. and {Mizuno}, T. and {Monte}, C. and {Monzani}, M.~E. and {Morselli}, A. and {Moskalenko}, I.~V. and {Murgia}, S. and {Naumann-Godo}, M. and {Nolan}, P.~L. and {Norris}, J.~P. and {Nuss}, E. and {Ohsugi}, T. and {Okumura}, A. and {Orlando}, E. and {Ormes}, J.~F. and {Ozaki}, M. and {Paneque}, D. and {Parent}, D. and {Pesce-Rollins}, M. and {Pierbattista}, M. and {Piron}, F. and {Pohl}, M. and {Prokhorov}, D. and {Rain{\`o}}, S. and {Rando}, R. and {Razzano}, M. and {Reposeur}, T. and {Ritz}, S. and {Parkinson}, P.~M. Saz and {Sgr{\`o}}, C. and {Siskind}, E.~J. and {Smith}, P.~D. and {Spinelli}, P. and {Strong}, A.~W. and {Takahashi}, H. and {Tanaka}, T. and {Thayer}, J.~G. and {Thayer}, J.~B. and {Thompson}, D.~J. and {Tibaldo}, L. and {Torres}, D.~F. and {Tosti}, G. and {Tramacere}, A. and {Troja}, E. and {Uchiyama}, Y. and {Vandenbroucke}, J. and {Vasileiou}, V. and {Vianello}, G. and {Vitale}, V. and {Waite}, A.~P. and {Wang}, P. and {Winer}, B.~L. and {Wood}, K.~S. and {Yang}, Z. and {Zimmer}, S. and {Bontemps}, S.},
        title = "{A Cocoon of Freshly Accelerated Cosmic Rays Detected by Fermi in the Cygnus Superbubble}",
      journal = {Science},
         year = 2011,
        month = nov,
       volume = {334},
       number = {6059},
        pages = {1103},
          doi = {10.1126/science.1210311},
       adsurl = {https://ui.adsabs.harvard.edu/abs/2011Sci...334.1103A}
}

@ARTICLE{Ab12,
       author = {{Abramowski}, A. and {Acero}, F. and {Aharonian}, F. and {Akhperjanian}, A.~G. and {Anton}, G. and {Balzer}, A. and {Barnacka}, A. and {Barres de Almeida}, U. and {Becherini}, Y. and {Becker}, J. and {Behera}, B. and {Bernl{\"o}hr}, K. and {Birsin}, E. and {Biteau}, J. and {Bochow}, A. and {Boisson}, C. and {Bolmont}, J. and {Bordas}, P. and {Brucker}, J. and {Brun}, F. and {Brun}, P. and {Bulik}, T. and {B{\"u}sching}, I. and {Carrigan}, S. and {Casanova}, S. and {Cerruti}, M. and {Chadwick}, P.~M. and {Charbonnier}, A. and {Chaves}, R.~C.~G. and {Cheesebrough}, A. and {Chounet}, L.-M. and {Clapson}, A.~C. and {Coignet}, G. and {Cologna}, G. and {Conrad}, J. and {Dalton}, M. and {Daniel}, M.~K. and {Davids}, I.~D. and {Degrange}, B. and {Deil}, C. and {Dickinson}, H.~J. and {Djannati-Ata{\"\i}}, A. and {Domainko}, W. and {O'C. Drury}, L. and {Dubois}, F. and {Dubus}, G. and {Dutson}, K. and {Dyks}, J. and {Dyrda}, M. and {Egberts}, K. and {Eger}, P. and {Espigat}, P. and {Fallon}, L. and {Farnier}, C. and {Fegan}, S. and {Feinstein}, F. and {Fernandes}, M.~V. and {Fiasson}, A. and {Fontaine}, G. and {F{\"o}rster}, A. and {F{\"u}{\ss}ling}, M. and {Gallant}, Y.~A. and {Gast}, H. and {G{\'e}rard}, L. and {Gerbig}, D. and {Giebels}, B. and {Glicenstein}, J.~F. and {Gl{\"u}ck}, B. and {Goret}, P. and {G{\"o}ring}, D. and {H{\"a}ffner}, S. and {Hague}, J.~D. and {Hampf}, D. and {Hauser}, M. and {Heinz}, S. and {Heinzelmann}, G. and {Henri}, G. and {Hermann}, G. and {Hinton}, J.~A. and {Hoffmann}, A. and {Hofmann}, W. and {Hofverberg}, P. and {Holler}, M. and {Horns}, D. and {Jacholkowska}, A. and {de Jager}, O.~C. and {Jahn}, C. and {Jamrozy}, M. and {Jung}, I. and {Kastendieck}, M.~A. and {Katarzy{\'n}ski}, K. and {Katz}, U. and {Kaufmann}, S. and {Keogh}, D. and {Khangulyan}, D. and {Kh{\'e}lifi}, B. and {Klochkov}, D. and {Klu{\.z}niak}, W. and {Kneiske}, T. and {Komin}, Nu. and {Kosack}, K. and {Kossakowski}, R. and {Laffon}, H. and {Lamanna}, G. and {Lennarz}, D. and {Lohse}, T. and {Lopatin}, A. and {Lu}, C.-C. and {Marandon}, V. and {Marcowith}, A. and {Masbou}, J. and {Maurin}, D. and {Maxted}, N. and {Mayer}, M. and {McComb}, T.~J.~L. and {Medina}, M.~C. and {M{\'e}hault}, J. and {Moderski}, R. and {Moulin}, E. and {Naumann}, C.~L. and {Naumann-Godo}, M. and {de Naurois}, M. and {Nedbal}, D. and {Nekrassov}, D. and {Nguyen}, N. and {Nicholas}, B. and {Niemiec}, J. and {Nolan}, S.~J. and {Ohm}, S. and {de O{\~n}a Wilhelmi}, E. and {Opitz}, B. and {Ostrowski}, M. and {Oya}, I. and {Panter}, M. and {Paz Arribas}, M. and {Pedaletti}, G. and {Pelletier}, G. and {Petrucci}, P.-O. and {Pita}, S. and {P{\"u}hlhofer}, G. and {Punch}, M. and {Quirrenbach}, A. and {Raue}, M. and {Rayner}, S.~M. and {Reimer}, A. and {Reimer}, O. and {Renaud}, M. and {de Los Reyes}, R. and {Rieger}, F. and {Ripken}, J. and {Rob}, L. and {Rosier-Lees}, S. and {Rowell}, G. and {Rudak}, B. and {Rulten}, C.~B. and {Ruppel}, J. and {Sahakian}, V. and {Sanchez}, D. and {Santangelo}, A. and {Schlickeiser}, R. and {Sch{\"o}ck}, F.~M. and {Schulz}, A. and {Schwanke}, U. and {Schwarzburg}, S. and {Schwemmer}, S. and {Sheidaei}, F. and {Sikora}, M. and {Skilton}, J.~L. and {Sol}, H. and {Spengler}, G. and {Stawarz}, {\L}. and {Steenkamp}, R. and {Stegmann}, C. and {Stinzing}, F. and {Stycz}, K. and {Sushch}, I. and {Szostek}, A. and {Tavernet}, J.-P. and {Terrier}, R. and {Tluczykont}, M. and {Valerius}, K. and {van Eldik}, C. and {Vasileiadis}, G. and {Venter}, C. and {Vialle}, J.~P. and {Viana}, A. and {Vincent}, P. and {V{\"o}lk}, H.~J. and {Volpe}, F. and {Vorobiov}, S. and {Vorster}, M. and {Wagner}, S.~J. and {Ward}, M. and {White}, R. and {Wierzcholska}, A. and {Zacharias}, M. and {Zajczyk}, A. and {Zdziarski}, A.~A. and {Zech}, A. and {Zechlin}, H.-S.},
        title = "{Discovery of extended VHE {\ensuremath{\gamma}}-ray emission from the vicinity of the young massive stellar cluster Westerlund 1}",
      journal = {\aap},
         year = 2012,
        month = jan,
       volume = {537},
          eid = {A114},
        pages = {A114},
          doi = {10.1051/0004-6361/201117928},
archivePrefix = {arXiv},
       eprint = {1111.2043},
 primaryClass = {astro-ph.HE},
       adsurl = {https://ui.adsabs.harvard.edu/abs/2012A&A...537A.114A}
}

@ARTICLE{hermes,
       author = {{Dundovic}, A. and {Evoli}, C. and {Gaggero}, D. and {Grasso}, D.},
        title = "{Simulating the Galactic multi-messenger emissions with HERMES}",
      journal = {\aap},
         year = 2021,
        month = sep,
       volume = {653},
          eid = {A18},
        pages = {A18},
          doi = {10.1051/0004-6361/202140801},
archivePrefix = {arXiv},
       eprint = {2105.13165},
 primaryClass = {astro-ph.HE},
       adsurl = {https://ui.adsabs.harvard.edu/abs/2021A&A...653A..18D}
}

@ARTICLE{h14pi,
       author = {{HI4PI Collaboration} and {Ben Bekhti}, N. and {Fl{\"o}er}, L. and {Keller}, R. and {Kerp}, J. and {Lenz}, D. and {Winkel}, B. and {Bailin}, J. and {Calabretta}, M.~R. and {Dedes}, L. and {Ford}, H.~A. and {Gibson}, B.~K. and {Haud}, U. and {Janowiecki}, S. and {Kalberla}, P.~M.~W. and {Lockman}, F.~J. and {McClure-Griffiths}, N.~M. and {Murphy}, T. and {Nakanishi}, H. and {Pisano}, D.~J. and {Staveley-Smith}, L.},
        title = "{HI4PI: A full-sky H I survey based on EBHIS and GASS}",
      journal = {\aap},
         year = 2016,
        month = oct,
       volume = {594},
          eid = {A116},
        pages = {A116},
          doi = {10.1051/0004-6361/201629178},
archivePrefix = {arXiv},
       eprint = {1610.06175},
 primaryClass = {astro-ph.GA},
       adsurl = {https://ui.adsabs.harvard.edu/abs/2016A&A...594A.116H}
}

@ARTICLE{dame,
       author = {{Dame}, T.~M. and {Hartmann}, Dap and {Thaddeus}, P.},
        title = "{The Milky Way in Molecular Clouds: A New Complete CO Survey}",
      journal = {\apj},
         year = 2001,
        month = feb,
       volume = {547},
       number = {2},
        pages = {792-813},
          doi = {10.1086/318388},
archivePrefix = {arXiv},
       eprint = {astro-ph/0009217},
 primaryClass = {astro-ph},
       adsurl = {https://ui.adsabs.harvard.edu/abs/2001ApJ...547..792D}
}

@ARTICLE{kelneraharonian,
       author = {{Kelner}, S.~R. and {Aharonian}, F.~A. and {Bugayov}, V.~V.},
        title = "{Energy spectra of gamma rays, electrons, and neutrinos produced at proton-proton interactions in the very high energy regime}",
      journal = {\prd},
         year = 2006,
        month = aug,
       volume = {74},
       number = {3},
          eid = {034018},
        pages = {034018},
          doi = {10.1103/PhysRevD.74.034018},
archivePrefix = {arXiv},
       eprint = {astro-ph/0606058},
 primaryClass = {astro-ph},
       adsurl = {https://ui.adsabs.harvard.edu/abs/2006PhRvD..74c4018K}
}

@ARTICLE{HAWCCrab2019,
       author = {{Abeysekara}, A.~U. and {Albert}, A. and {Alfaro}, R. and {Alvarez}, C. and {{\'A}lvarez}, J.~D. and {Camacho}, J.~R. Angeles and {Arceo}, R. and {Arteaga-Vel{\'a}zquez}, J.~C. and {Arunbabu}, K.~P. and {Avila Rojas}, D. and {Ayala Solares}, H.~A. and {Baghmanyan}, V. and {Belmont-Moreno}, E. and {BenZvi}, S.~Y. and {Brisbois}, C. and {Caballero-Mora}, K.~S. and {Capistr{\'a}n}, T. and {Carrami{\~n}ana}, A. and {Casanova}, S. and {Cotti}, U. and {Cotzomi}, J. and {Couti{\~n}o de Le{\'o}n}, S. and {De la Fuente}, E. and {de Le{\'o}n}, C. and {Dichiara}, S. and {Dingus}, B.~L. and {DuVernois}, M.~A. and {D{\'\i}az-V{\'e}lez}, J.~C. and {Ellsworth}, R.~W. and {Engel}, K. and {Espinoza}, C. and {Fick}, B. and {Fleischhack}, H. and {Fraija}, N. and {Galv{\'a}n-G{\'a}mez}, A. and {Garc{\'\i}a-Gonz{\'a}lez}, J.~A. and {Garfias}, F. and {Gonz{\'a}lez}, M.~M. and {Goodman}, J.~A. and {Harding}, J.~P. and {Hernandez}, S. and {Hinton}, J. and {Hona}, B. and {Hueyotl-Zahuantitla}, F. and {Hui}, C.~M. and {H{\"u}ntemeyer}, P. and {Iriarte}, A. and {Jardin-Blicq}, A. and {Joshi}, V. and {Kaufmann}, S. and {Kieda}, D. and {Lara}, A. and {Lee}, W.~H. and {Le{\'o}n Vargas}, H. and {Linnemann}, J.~T. and {Longinotti}, A.~L. and {Luis-Raya}, G. and {Lundeen}, J. and {Malone}, K. and {Marinelli}, S.~S. and {Martinez}, O. and {Martinez-Castellanos}, I. and {Mart{\'\i}nez-Castro}, J. and {Mart{\'\i}nez-Huerta}, H. and {Matthews}, J.~A. and {Miranda-Romagnoli}, P. and {Morales-Soto}, J.~A. and {Moreno}, E. and {Mostaf{\'a}}, M. and {Nayerhoda}, A. and {Nellen}, L. and {Newbold}, M. and {Nisa}, M.~U. and {Noriega-Papaqui}, R. and {Peisker}, A. and {P{\'e}rez-P{\'e}rez}, E.~G. and {Pretz}, J. and {Ren}, Z. and {Rho}, C.~D. and {Rivi{\`e}re}, C. and {Rosa-Gonz{\'a}lez}, D. and {Rosenberg}, M. and {Ruiz-Velasco}, E. and {Salazar}, H. and {Salesa Greus}, F. and {Sandoval}, A. and {Schneider}, M. and {Schoorlemmer}, H. and {Seglar Arroyo}, M. and {Sinnis}, G. and {Smith}, A.~J. and {Springer}, R.~W. and {Surajbali}, P. and {Tabachnick}, E. and {Tanner}, M. and {Tibolla}, O. and {Tollefson}, K. and {Torres}, I. and {Weisgarber}, T. and {Westerhoff}, S. and {Wood}, J. and {Yapici}, T. and {Zepeda}, A. and {Zhou}, H. and {HAWC Collaboration}},
        title = "{Measurement of the Crab Nebula Spectrum Past 100 TeV with HAWC}",
      journal = {\apj},
         year = 2019,
        month = aug,
       volume = {881},
       number = {2},
          eid = {134},
        pages = {134},
          doi = {10.3847/1538-4357/ab2f7d},
archivePrefix = {arXiv},
       eprint = {1905.12518},
 primaryClass = {astro-ph.HE},
       adsurl = {https://ui.adsabs.harvard.edu/abs/2019ApJ...881..134A}
}

@ARTICLE{HAWCML,
       author = {{Alfaro}, R. and {Alvarez}, C. and {Andr{\'e}s}, A. and {Anita-Rangel}, E. and {Araya}, M. and {Arteaga-Vel{\'a}zquez}, J.~C. and {Avila Rojas}, D. and {Ayala Solares}, H.~A. and {Babu}, R. and {Bangale}, P. and et al.},
        title = "{HAWC Performance Enhanced by Machine Learning in Gamma-hadron Separation}",
      journal = {\apj},
         year = 2025,
        month = oct,
       volume = {992},
       number = {1},
          eid = {156},
        pages = {156},
          doi = {10.3847/1538-4357/ae0186},
archivePrefix = {arXiv},
       eprint = {2506.18277},
 primaryClass = {astro-ph.IM},
       adsurl = {https://ui.adsabs.harvard.edu/abs/2025ApJ...992..156A}
}

@article{HAWCP5,
    author = "Albert, A. . and others",
    collaboration = "HAWC",
    title = "{Performance of the HAWC Observatory and TeV Gamma-Ray Measurements of the Crab Nebula with Improved Extensive Air Shower Reconstruction Algorithms}",
    eprint = "2405.06050",
    archivePrefix = "arXiv",
    primaryClass = "astro-ph.HE",
    doi = "10.3847/1538-4357/ad5f2d",
    journal = "Astrophys. J.",
    volume = "972",
    number = "2",
    pages = "144",
    year = "2024"
}

@ARTICLE{threeML,
       author = {{Vianello}, Giacomo and {Lauer}, Robert J. and {Younk}, Patrick and {Tibaldo}, Luigi and {Burgess}, James M. and {Ayala}, Hugo and {Harding}, Patrick and {Hui}, Michelle and {Omodei}, Nicola and {Zhou}, Hao},
        title = "{The Multi-Mission Maximum Likelihood framework (3ML)}",
      journal = {arXiv e-prints},
         year = 2015,
        month = jul,
          eid = {arXiv:1507.08343},
        pages = {arXiv:1507.08343},
          doi = {10.48550/arXiv.1507.08343},
archivePrefix = {arXiv},
       eprint = {1507.08343},
 primaryClass = {astro-ph.HE},
       adsurl = {https://ui.adsabs.harvard.edu/abs/2015arXiv150708343V}
}

@article{hawchal,
    author = "Abeysekara, Anushka Udara and others",
    collaboration = "HAWC",
    title = "{Characterizing gamma-ray sources with HAL (HAWC Accelerated likelihood) and 3ML}",
    doi = "10.22323/1.395.0828",
    journal = "PoS",
    volume = "ICRC2021",
    pages = "828",
    year = "2021"
}

@ARTICLE{hawcimage,
       author = {{Babu}, Rishi and {Wentworth}, Palmer and {Herzog}, Ian and {Salazar}, Dan and {Nisa}, Mehr Un},
        title = "{Improving Gamma-ray Source Search with Image Processing}",
      journal = {arXiv e-prints},
         year = 2025,
        month = jul,
          eid = {arXiv:2507.10307},
        pages = {arXiv:2507.10307},
          doi = {10.48550/arXiv.2507.10307},
archivePrefix = {arXiv},
       eprint = {2507.10307},
 primaryClass = {astro-ph.HE},
       adsurl = {https://ui.adsabs.harvard.edu/abs/2025arXiv250710307B}
}

@article{bic,
author = {Gideon Schwarz},
title = {{Estimating the Dimension of a Model}},
volume = {6},
journal = {The Annals of Statistics},
number = {2},
publisher = {Institute of Mathematical Statistics},
pages = {461 -- 464},
year = {1978},
doi = {10.1214/aos/1176344136},
URL = {https://doi.org/10.1214/aos/1176344136}
}

@ARTICLE{aic,
  author={Akaike, H.},
  journal={IEEE Transactions on Automatic Control}, 
  title={A new look at the statistical model identification}, 
  year={1974},
  volume={19},
  number={6},
  pages={716-723},
  doi={10.1109/TAC.1974.1100705}}

@article{Blasi2013,
  author  = {Blasi, Pasquale},
  title   = {The Origin of Galactic Cosmic Rays},
  journal = {Astronomy and Astrophysics Review},
  volume  = {21},
  pages   = {70},
  year    = {2013},
  doi     = {10.1007/s00159-013-0070-7}
}

@article{Ptuskin2006,
  author  = {Ptuskin, V. S. and Moskalenko, I. V. and Jones, F. C. and Strong, A. W. and Zirakashvili, V. N.},
  title   = {Dissipation of Magnetohydrodynamic Waves on Energetic Particles: Impact on Interstellar Turbulence and Cosmic-Ray Transport},
  journal = {The Astrophysical Journal},
  volume  = {642},
  pages   = {902--916},
  year    = {2006},
  doi     = {10.1086/501117}
}

@article{Atoyan1995,
  author  = {Atoyan, A. M. and Aharonian, F. A. and V{\"o}lk, H. J.},
  title   = {Electrons and positrons in the Galactic cosmic rays},
  journal = {Physical Review D},
  volume  = {52},
  pages   = {3265--3275},
  year    = {1995},
  doi     = {10.1103/PhysRevD.52.3265}
}

@article{Ohira2011,
    author = {Ohira, Yutaka and Murase, Kohta and Yamazaki, Ryo},
    title = {Gamma-rays from molecular clouds illuminated by cosmic rays escaping from interacting supernova remnants},
    journal = {Monthly Notices of the Royal Astronomical Society},
    volume = {410},
    number = {3},
    pages = {1577-1582},
    year = {2011},
    month = {01},
    issn = {0035-8711},
    doi = {10.1111/j.1365-2966.2010.17539.x},
    url = {https://doi.org/10.1111/j.1365-2966.2010.17539.x},
    eprint = {https://academic.oup.com/mnras/article-pdf/410/3/1577/2849859/mnras0410-1577.pdf},
}

@INPROCEEDINGS{1982BAAS...14..925S,
       author = {{Simon}, M. and {Fischer}, J.},
        title = "{Velocity Resolved IR Spectroscopy of S106 IRS 4}",
    booktitle = {Bulletin of the American Astronomical Society},
         year = 1982,
       volume = {14},
        month = sep,
        pages = {925},
       adsurl = {https://ui.adsabs.harvard.edu/abs/1982BAAS...14..925S}
}

@ARTICLE{1983ApJ...272..154B,
       author = {{Bally}, J. and {Snell}, R.~L. and {Predmore}, R.},
        title = "{Radio images of the bipolar HII region S 106.}",
      journal = {\apj},
         year = 1983,
        month = sep,
       volume = {272},
        pages = {154-162},
          doi = {10.1086/161272},
       adsurl = {https://ui.adsabs.harvard.edu/abs/1983ApJ...272..154B}
}

@ARTICLE{2006AJ....131.1608O,
       author = {{Oasa}, Yumiko and {Tamura}, Motohide and {Nakajima}, Yasushi and {Itoh}, Yoichi and {Maihara}, Toshinori and {Iwamuro}, Fumihiko and {Motohara}, Kentaro and {Hayashi}, Saeko S. and {Hayashi}, Masahiko and {Kaifu}, Norio},
        title = "{Very Low Luminosity Young Cluster and the Luminosity and Mass Functions in S106}",
      journal = {\aj},
         year = 2006,
        month = mar,
       volume = {131},
       number = {3},
        pages = {1608-1628},
          doi = {10.1086/500086},
       adsurl = {https://ui.adsabs.harvard.edu/abs/2006AJ....131.1608O}
}

@article{Bally_2022,
doi = {10.3847/1538-4357/ac30de},
url = {https://doi.org/10.3847/1538-4357/ac30de},
year = {2022},
month = {jan},
publisher = {The American Astronomical Society},
volume = {924},
number = {2},
pages = {50},
author = {Bally, John and Chia, Zen and Ginsburg, Adam and Reipurth, Bo and Tanaka, Kei E. I and Zinnecker, Hans and Faulhaber, John},
title = {Supersonic Expansion of the Bipolar H ii Region Sh2-106: A 3500 Year Old Explosion?},
journal = {The Astrophysical Journal}
}

@ARTICLE{2026JHEAp..5000465B,
       author = {{Bonollo}, A. and {Esposito}, P. and {Giuliani}, A. and {Caraveo}, P. and {Galanti}, G. and {Crestan}, S. and {Rigoselli}, M. and {Mereghetti}, S.},
        title = "{Morphology of young massive stellar clusters with next-generation IACTs}",
      journal = {Journal of High Energy Astrophysics},
         year = 2026,
        month = feb,
       volume = {50},
          eid = {100465},
        pages = {100465},
          doi = {10.1016/j.jheap.2025.100465},
archivePrefix = {arXiv},
       eprint = {2509.20150},
 primaryClass = {astro-ph.HE},
       adsurl = {https://ui.adsabs.harvard.edu/abs/2026JHEAp..5000465B}
}

@ARTICLE{hawcj2019,
doi = {10.3847/1538-4357/abecda},
url = {https://doi.org/10.3847/1538-4357/abecda},
year = {2021},
month = {apr},
publisher = {The American Astronomical Society},
volume = {911},
number = {2},
pages = {143},
author = {Albert, A. and Alfaro, R. and Alvarez, C. and Arteaga-Velázquez, J. C. and Arunbabu, K. P. and Avila Rojas, D. and Ayala Solares, H. A. and Baghmanyan, V. and Belmont-Moreno, E. and Brisbois, C. and Caballero-Mora, K. S. and Capistrán, T. and Carramiñana, A. and Casanova, S. and Cotzomi, J. and Coutiño de León, S. and De la Fuente, E. and Diaz Hernandez, R. and Dingus, B. L. and DuVernois, M. A. and Durocher, M. and Engel, K. and Espinoza, C. and Fraija, N. and Garcia, D. and García-González, J. A. and Giacinti, G. and González, M. M. and Goodman, J. A. and Harding, J. P. and Hinton, J. and Hona, B. and Huang, D. and Hueyotl-Zahuantitla, F. and Huentemeyer, P. and Jardin-Blicq, A. and Joshi, V. and Lee, W. H. and León Vargas, H. and Linnemann, J. T. and Longinotti, A. L. and Luis-Raya, G. and Lundeen, J. and López-Coto, R. and Malone, K. and Martinez, O. and Martínez-Castro, J. and Matthews, J. A. and Miranda-Romagnoli, P. and Morales-Soto, J. A. and Moreno, E. and Mostafá, M. and Nayerhoda, A. and Nellen, L. and Newbold, M. and Nisa, M. U. and Noriega-Papaqui, R. and Olivera-Nieto, L. and Omodei, N. and Peisker, A. and Pérez Araujo, Y. and Pérez-Pérez, E. G. and Rho, C. D. and Rosa-González, D. and Ruiz-Velasco, E. and Salazar, H. and Salesa Greus, F. and Sandoval, A. and Schneider, M. and Schoorlemmer, H. and Serna-Franco, J. and Smith, A. J. and Springer, R. W. and Surajbali, P. and Tollefson, K. and Torres, I. and Turner, R. and Ureña-Mena, F. and Weisgarber, T. and Willox, E. and Zhou, H. and de León, C. and HAWC Collaboration},
title = {Spectrum and Morphology of the Very-high-energy Source HAWC J2019+368},
journal = {The Astrophysical Journal}
}

@ARTICLE{hawcic443,
       author = {{Alfaro}, R. and {Alvarez}, C. and {Araya}, M. and {Arteaga-Vel{\'a}zquez}, J.~C. and {Avila Rojas}, D. and {Ayala Solares}, H.~A. and {Babu}, R. and {Bangale}, P. and {Bernal}, A. and {Caballero-Mora}, K.~S. and {Capistran}, T. and {Carrami{\~n}ana}, A. and {Casanova}, S. and {Cotti}, U. and {Cotzomi}, J. and {Couti{\~n}o de Le{\'o}n}, S. and {De la Fuente}, E. and {Depaoli}, D. and {Desiati}, P. and {Di Lalla}, N. and {Diaz Hernandez}, R. and {Dingus}, B.~L. and {DuVernois}, M.~A. and {D{\'\i}az-V{\'e}lez}, J.~C. and {Ergin}, T. and {Espinoza}, C. and {Fang}, K. and {Fraija}, N. and {Fraija}, S. and {Garc{\'\i}a-Gonz{\'a}lez}, J.~A. and {Goksu}, H. and {Gonz{\'a}lez-Cervera}, J.~A. and {Gonz{\'a}lez}, M.~M. and {Goodman}, J.~A. and {Groetsch}, S. and {Harding}, J.~P. and {Hern{\'a}ndez-Cadena}, S. and {Herzog}, I. and {Hinton}, J. and {Huang}, D. and {Hueyotl-Zahuantitla}, F. and {H{\"u}ntemeyer}, P. and {Kaufmann}, S. and {Lara}, A. and {Lee}, J. and {Le{\'o}n Vargas}, H. and {Linnemann}, J.~T. and {Longinotti}, A.~L. and {Luis-Raya}, G. and {Malone}, K. and {Martinez}, O. and {Mart{\'\i}nez-Castro}, J. and {Matthews}, J.~A. and {Miranda-Romagnoli}, P. and {Montes}, J.~A. and {Morales-Soto}, J.~A. and {Moreno}, E. and {Mostaf{\'a}}, M. and {Najafi}, M. and {Nellen}, L. and {Nisa}, M.~U. and {Noriega-Papaqui}, R. and {Olivera-Nieto}, L. and {Omodei}, N. and {Osorio}, M. and {Ponce}, E. and {P{\'e}rez Araujo}, Y. and {P{\'e}rez-P{\'e}rez}, E.~G. and {Rho}, C.~D. and {Rosa-Gonz{\'a}lez}, D. and {Roth}, M. and {Ruiz-Velasco}, E. and {Salazar}, H. and {Sandoval}, A. and {Schneider}, M. and {Schwefer}, G. and {Serna-Franco}, J. and {Smith}, A.~J. and {Son}, Y. and {Springer}, R.~W. and {Tibolla}, O. and {Tollefson}, K. and {Torres}, I. and {Torres-Escobedo}, R. and {Turner}, R. and {Wang}, X. and {Wang}, Z. and {Watson}, I.~J. and {Wu}, H. and {Yu}, S. and {Yun-C{\'a}rcamo}, S. and {Zhou}, H. and {de Le{\'o}n}, C. and {HAWC Collaboration}},
        title = "{Study of the IC 443 Region with the HAWC Observatory}",
      journal = {\apj},
         year = 2025,
        month = oct,
       volume = {992},
       number = {1},
          eid = {22},
        pages = {22},
          doi = {10.3847/1538-4357/adf7aa},
archivePrefix = {arXiv},
       eprint = {2501.12613},
 primaryClass = {astro-ph.HE},
       adsurl = {https://ui.adsabs.harvard.edu/abs/2025ApJ...992...22A}
}

@ARTICLE{kinematicdistancetool,
       author = {{Wenger}, Trey V. and {Balser}, Dana S. and {Anderson}, L.~D. and {Bania}, T.~M.},
        title = "{Kinematic Distances: A Monte Carlo Method}",
      journal = {\apj},
         year = 2018,
        month = mar,
       volume = {856},
       number = {1},
          eid = {52},
        pages = {52},
          doi = {10.3847/1538-4357/aaaec8},
archivePrefix = {arXiv},
       eprint = {1802.04203},
 primaryClass = {astro-ph.GA},
       adsurl = {https://ui.adsabs.harvard.edu/abs/2018ApJ...856...52W}
}

@article{LHCygnusC2024,
title = {An ultrahigh-energy γ-ray bubble powered by a super PeVatron},
journal = {Science Bulletin},
volume = {69},
number = {4},
pages = {449-457},
year = {2024},
issn = {2095-9273},
doi = {https://doi.org/10.1016/j.scib.2023.12.040},
url = {https://www.sciencedirect.com/science/article/pii/S209592732300912X},
author = { {LHAASO Collaboration}},
}

@article{moderski,
    author = {Moderski, Rafal and Sikora, Marek and Coppi, Paolo S. and Aharonian, Felix},
    title = {Klein—Nishina effects in the spectra of non-thermal sources immersed in external radiation fields},
    journal = {Monthly Notices of the Royal Astronomical Society},
    volume = {363},
    number = {3},
    pages = {954-966},
    year = {2005},
    month = {11},
    issn = {0035-8711},
    doi = {10.1111/j.1365-2966.2005.09494.x},
    url = {https://doi.org/10.1111/j.1365-2966.2005.09494.x},
    eprint = {https://academic.oup.com/mnras/article-pdf/363/3/954/3954953/363-3-954.pdf},
}

@article{adams2015,
  title={SOFIA/FORCAST Observations of Warm Dust in S106: A Fragmented Environment},
  author={Adams, JD and Herter, TL and Hora, JL and Schneider, N and Lau, RM and Staguhn, JG and Simon, R and Smith, N and Gehrz, RD and Allen, LE and others},
  journal={The Astrophysical Journal},
  volume={814},
  number={1},
  pages={54},
  year={2015},
  publisher={The American Astronomical Society}
}

@ARTICLE{2020ApJ...905...76A,
       author = {{Albert}, A. and {Alfaro}, R. and {Alvarez}, C. and {Camacho}, J.~R. Angeles and {Arteaga-Vel{\'a}zquez}, J.~C. and {Arunbabu}, K.~P. and {Avila Rojas}, D. and {Ayala Solares}, H.~A. and {Baghmanyan}, V. and {Belmont-Moreno}, E. and {BenZvi}, S.~Y. and {Brisbois}, C. and {Caballero-Mora}, K.~S. and {Capistr{\'a}n}, T. and {Carrami{\~n}ana}, A. and {Casanova}, S. and {Cotti}, U. and {Couti{\~n}o de Le{\'o}n}, S. and {De la Fuente}, E. and {Diaz Hernandez}, R. and {Diaz-Cruz}, L. and {Dingus}, B.~L. and {DuVernois}, M.~A. and {Durocher}, M. and {D{\'\i}az-V{\'e}lez}, J.~C. and {Ellsworth}, R.~W. and {Engel}, K. and {Espinoza}, C. and {Fan}, K.~L. and {Fang}, K. and {Alonso}, M. Fern{\'a}ndez and {Fleischhack}, H. and {Fraija}, N. and {Galv{\'a}n-G{\'a}mez}, A. and {Garcia}, D. and {Garc{\'\i}a-Gonz{\'a}lez}, J.~A. and {Garfias}, F. and {Giacinti}, G. and {Gonz{\'a}lez}, M.~M. and {Goodman}, J.~A. and {Harding}, J.~P. and {Hernandez}, S. and {Hinton}, J. and {Hona}, B. and {Huang}, D. and {Hueyotl-Zahuantitla}, F. and {H{\"u}ntemeyer}, P. and {Iriarte}, A. and {Jardin-Blicq}, A. and {Joshi}, V. and {Kieda}, D. and {Lara}, A. and {Lee}, W.~H. and {Le{\'o}n Vargas}, H. and {Linnemann}, J.~T. and {Longinotti}, A.~L. and {Luis-Raya}, G. and {Lundeen}, J. and {L{\'o}pez-Coto}, R. and {Malone}, K. and {Marandon}, V. and {Martinez}, O. and {Martinez-Castellanos}, I. and {Mart{\'\i}nez-Castro}, J. and {Matthews}, J.~A. and {Miranda-Romagnoli}, P. and {Morales-Soto}, J.~A. and {Moreno}, E. and {Mostaf{\'a}}, M. and {Nayerhoda}, A. and {Nellen}, L. and {Newbold}, M. and {Nisa}, M.~U. and {Noriega-Papaqui}, R. and {Olivera-Nieto}, L. and {Omodei}, N. and {Peisker}, A. and {P{\'e}rez Araujo}, Y. and {P{\'e}rez-P{\'e}rez}, E.~G. and {Ren}, Z. and {Rho}, C.~D. and {Rivi{\`e}re}, C. and {Rosa-Gonz{\'a}lez}, D. and {Ruiz-Velasco}, E. and {Salazar}, H. and {Salesa Greus}, F. and {Sandoval}, A. and {Schneider}, M. and {Schoorlemmer}, H. and {Serna}, F. and {Sinnis}, G. and {Smith}, A.~J. and {Springer}, R.~W. and {Surajbali}, P. and {Tollefson}, K. and {Torres}, I. and {Torres-Escobedo}, R. and {Ukwatta}, T.~N. and {Ure{\~n}a-Mena}, F. and {Weisgarber}, T. and {Werner}, F. and {Willox}, E. and {Zepeda}, A. and {Zhou}, H. and {de Le{\'o}n}, C. and {{\'A}lvarez}, J.~D. and {HAWC Collaboration}},
        title = "{3HWC: The Third HAWC Catalog of Very-high-energy Gamma-Ray Sources}",
      journal = {\apj},
         year = 2020,
        month = dec,
       volume = {905},
       number = {1},
          eid = {76},
        pages = {76},
          doi = {10.3847/1538-4357/abc2d8},
archivePrefix = {arXiv},
       eprint = {2007.08582},
 primaryClass = {astro-ph.HE},
       adsurl = {https://ui.adsabs.harvard.edu/abs/2020ApJ...905...76A}
}
\end{document}